\documentclass[a4paper,11pt]{article}
\usepackage{amssymb,amsmath,framed,color,mathrsfs}
\usepackage[colorlinks]{hyperref}
\usepackage{graphicx}

\newcommand{\rem}[1]{}

\newcommand{\de}{{\rm d}}

\def\contract{\makebox[1.2em][c]{\mbox{\rule{.6em}
{.01truein}\rule{.01truein}{.6em}}}}

\newcommand{\comment}[1]{\vspace{1 mm}\par
\marginpar{\large\underline{}}\noindent
\framebox{\begin{minipage}[c]{0.982 \textwidth} \tt #1
\end{minipage}}\vspace{2 mm}\par}

\def\contract{\makebox[1.2em][c]{\mbox{\rule{.6em}
{.01truein}\rule{.01truein}{.6em}}}}

\newtheorem{remark}{Remark}

\title{Hybrid models for complex fluids with multipolar interactions\vspace{-.4cm}}

\author{Cesare Tronci\\\footnotesize\it Section de Math\'ematiques, \'Ecole
Polytechnique F\'ed\'erale de Lausanne, Switzerland\\
\footnotesize\it Department of Mathematics, University of Surrey, Guildford, United Kingdom}
\date{\small\sf For Tudor Ratiu, on the occasion of his 60th birthday\vspace{-.2cm}}

\begin{document}
\maketitle

\begin{abstract}
Multipolar order in complex fluids is described by statistical correlations. This paper presents a novel dynamical approach, which accounts for microscopic effects on the order parameter space. Indeed, the order parameter field is replaced by a statistical distribution function that is carried by the fluid flow. Inspired by Doi's model of colloidal suspensions, the present theory is derived from a \emph{hybrid} moment closure for Yang-Mills Vlasov plasmas. {\color{black}This hybrid formulation is constructed under the assumption that inertial effects dominate over dissipative phenomena, so that the total energy is conserved.} After presenting the basic geometric properties of the theory, the effect of Yang-Mills fields is considered and a direct application is presented to magnetized fluids with quadrupolar order (\emph{spin nematic} phases). Hybrid models are also formulated for complex fluids with symmetry breaking. For the special case of liquid crystals, the moment method can be applied to the hybrid formulation to study to the dynamics of cubatic phases.
\vspace{-.2cm}
\end{abstract}

\tableofcontents

\section{Introduction}

\subsection{Micromotion and its symmetry properties}
Complex fluids are fluid systems whose particles carry an internal structure. Interesting examples are found in the dynamics of ferromagnetic fluids and spin glasses \cite{Fischer,HaSa}. In this case, the particles carry a spin magnetic moment undergoing its own rotational motion (\emph{micromotion}), usually under the influence of external magnetic fields. This micromotion exhibits a full rotational symmetry so that the corresponding dynamics is entirely expressed in terms of the spin angular momentum variable, a situation also occurring for the rotational motion of a free rigid body on the rotation Lie group $SO(3)$. However, this fully symmetric configuration differs from many situations occurring in condensed matter physics, which exhibit the symmetry breaking phenomenon. The most celebrated example is provided by the dynamics of nematic molecules, which are endowed with a special orientation identified with an unsigned unit vector ${\bf n}\sim -{\bf n}$ ({\it director}). The presence of this special direction is responsible for symmetry breaking: the micromotion is no more symmetric under all possible $SO(3)$ rotations in physical space, but rather the molecule exhibits only an $O(2)$ residual symmetry. Then, the dynamics of the nematic molecule is described in terms of two variables: its director and the associated angular momentum.

\subsection{Correlation effects: complex media with multipolar order}
In common continuum descriptions of complex fluids, the effects of micromotion are introduced through a macroscopic continuum field taking values in the space $M$ of internal dynamical variables (e.g. the spin or the director). This space is usually known as \emph{order parameter space}. More particularly, one introduces a map $\phi:\Bbb{R}^3\to M$ (\emph{order parameter field}) from physical space $\Bbb{R}^3$ to the order parameter space $M$. This quantity is assumed to arise from an averaging process involving an appropriate statistical distribution function $\rho\in C^\infty(\Bbb{R}^3\times M)$. While this averaging process has been widely used in equilibrium statistical approaches since Onsager's work \cite{Onsager}, its use in the study of dynamical aspects has always been difficult. As a result, microscopic dynamical properties of orientational order are neglected in many complex fluid models.

On the other hand, an increasing number of complex media are being discovered in which statistical correlation effects determine the particle interaction. For example, magnets exhibiting quadrupolar order (spin correlations) known as ``spin nematics'' have been investigated since the early 80's  \cite{AnGr} and their dynamical description remains an open question in soft matter modeling \cite{TsMi2006}. These systems possess an experimental realization, which was reported in \cite{LaDoLhSiTr}. Recently, investigations have begun that study magnetic media with multipolar order higher than two: in \cite{SuLuLa} a phase diagram was presented to identify quadrupolar, octupolar and hexadecupolar ferromagnetic phases. Higher-order correlation effects also play some role in the dynamics of  liquid crystals. Indeed, when molecules possess a special shape (e.g. hard-cut spheres), the emergence of a ``cubatic phase'' in liquid crystals \cite{VeFr,DuDeMaWi} has been observed in computer simulations for about 20 years. This phase owes its name to the cubic orientational order (third-order director correlations) of its molecular interaction. The word `cubatic' is used to distinguish this phase from cubic phases, which possess also translational order. Cubatic phases have been shown to emerge for a variety of molecular shapes, see e.g. \cite{BaStTo} for how cubatic phases emerge in cubelike molecules.

The effects of multipolar order are usually studied by introducing a {\color{black}tensor} order parameter field, which {\color{black}encodes} the correlation effects. For example, this is the approach adopted for { cubatic liquid crystals \cite{DuDeMaWi} and magnets with quadrupolar order \cite{Isayev}}. The question whether such a description is satisfactory {\color{black}may require} complicated  kinetic theory arguments based on BBGKY-like hierarchies. This is precisely the case for Doi's celebrated description of colloidal solutions \cite{DoEd1988}. {\color{black}Doi's} theory aims to couple the dynamics of an ordinary fluid with a Smoluchowski kinetic equation for the dynamics of colloidal particles with orientation. In this way, the fluid correlation effects are encoded as usual in the equation of state, while the statistical information on the colloidal particles is carried along by the fluid through the Smoluchowski distribution function. This description stands as the main inspiration for the present work, whose purpose is to 
introduce a unified systematic framework for the hybrid fluid-kinetic description of perfect complex fluids, by invoking well established Lie symmetry principles. This framework can be used to model the fluid stress tensor, which remains an outstanding open question for these systems. An important contribution is due to Constantin \cite{Co2005}, who derived a general form of the stress tensor from energy principle considerations. {\color{black}However, contrarily to Doi theory, the present treatment considers inertial effects that dominate over dissipation, so that the total energy is conserved.}

Besides Doi's approach \cite{DoEd1988}, other attempts to account for correlation effects include mesoscopic theories of liquid crystals \cite{BlEhMu91,BlEhMu92}. These theories involve a set of fluid balance equations on the Cartesian
product $\Bbb{R}^3\times M$ of physical space and order parameter space. However, these equations are enormously difficult to handle since each different fluid variable (e.g. density and momentum) is defined on a space with a high number of dimensions.

\rem{ %%%%%%%%%%%%%%%%%%%%%%%%%%%%%%%%
Besides the celebrated kinetic approach of Doi and Edwards \cite{DoEd1988}, an interesting attempt to include microscopic effects is the \emph{mesoscopic approach} for liquid crystals \cite{BlEhMu91,BlEhMu92}. This approach leads to a theory of fluid balance equations on the Cartesian
product $\Bbb{R}^3\times M$ of physical space and order parameter space. In this theory one starts with a time-dependent probability distribution function $f\in C^\infty(T^*\Bbb{R}^3\times T^*M)$ on the extended phase space $\Bbb{R}^6\times T^*M$ and later averages over the momentum coordinates to derive the time evolution of mesoscopic quantities depending on the position coordinate in  $\Bbb{R}^3\times M$.
However, despite their rigorous validity, the mesoscopic equations are enormously difficult to handle since they involve different fluid variables (e.g. density and momentum) each of them defined on a space with a high number of dimensions, i.e. $4+\operatorname{dim}M$ including time.
}   %%%%%%%%%%%%%%%%%%%%%%%%%%%%%%%%

Additional considerations are needed when orientational defects (i.e. disclinations) are present in the system. In this case, the interaction between fluid, order parameter variables and defect dynamics is of central interest in modern soft matter physics. To the author's knowledge, no microscopic description of this interaction is available in the literature. This description is one of the aims of the present paper.

{\color{black}Before introducing the hybrid approach to multipolar interactions, the next section reviews the Yang-Mills-Vlasov system \cite{GiHoKu1983} governing, e.g. quark-gluon plasmas. Its fluid closure (in particular the MHD limit \cite{HoKu1984}) provides the geometric basis of many complex fluid models that inspired the present work.} 

\subsection{Yang-Mills Vlasov plasmas: a review}

The kinetic theory of Yang-Mills plasmas finds its roots in the definition of a distribution function $f(\mathbf{x,p},\sigma,t)$, where $(\mathbf{x,p})\in T^*\Bbb{R}^3=\Bbb{R}^6$ are the usual position-momentum coordinates, while $\sigma\in\mathfrak{o}^*$ is the single-particle Yang-Mills charge belonging to the dual space of some Lie algebra $\mathfrak{o}$. When the standard Klimontovich method of kinetic theory \cite{Kl1982} is applied { in} this setting, one finds the mean-field kinetic equation (Yang-Mills Vlasov, or YM Vlasov) \cite{GiHoKu1983,MoMaRa84}
\begin{equation}\label{YMV}
\frac{\partial f}{\partial t}+\left\{f,\frac{\delta H}{\delta
f}\right\}+\left\langle\sigma,\left[\frac{\partial f}{\partial
\sigma},\frac{\partial}{\partial \sigma}\frac{\delta H}{\delta
f}\right]\right\rangle=0
\end{equation}
where $\{\cdot,\cdot\}$ denotes the canonical Poisson bracket on $\Bbb{R}^3\times\Bbb{R}^3$ and $\left[\cdot,\cdot\right]$ denotes the Lie bracket on $\mathfrak{o}$, while $\langle\cdot,\cdot\rangle$ is the pairing between $\mathfrak{o}$ and its dual.
%\begin{equation}
%\{h,g\}_1:=\{h,g\}+\left\langle\sigma,\left[\frac{\partial h}{\partial
%\sigma},\frac{\partial g}{\partial \sigma}\right]\right\rangle
%\end{equation}
This equation is coupled to the Yang-Mills field equations to form the Yang-Mills Vlasov system, with Hamiltonian
\begin{equation}
H=\frac12 \int f\left|\mathbf{p}-\left\langle\sigma,\mathbf{A}\right\rangle\right|^2
d\mathbf{x}\,d \mathbf{p}\,d\sigma  - \int f \left\langle\sigma,A_0\right\rangle d\mathbf{x}\,d
\mathbf{p}\,d\sigma
+\frac12\int\left( \left|d^\mathbf{A} \mathbf{A}\right|^2+\left|\mathbf{E}\right|^2\right)d\mathbf{x}
{+}\int\left\langle \mathbf{E},d^\mathbf{A}A_0\right\rangle d\mathbf{x} 
\label{YMVHam}
\end{equation}
where $(\mathbf{A,E})\in
\Omega^1(\mathbb{R}^3,\mathfrak{o})\times\Omega^2(\mathbb{R}^3,\mathfrak{o}^*)$
are the YM field variables consisting of the vector potential $\mathbf{A}$
and the electric field part $\mathbf{E}$. Here
$\Omega^1(\mathbb{R}^3,\mathfrak{o})$ and
$\Omega^2(\mathbb{R}^3,\mathfrak{o}^*)$ are $\mathfrak{o}$-valued
one-forms and $\mathfrak{o}^*$-valued two-forms respectively
($\Omega^n(M)$ denotes a generic $n$-form on the manifold $M$),
while ${\color{black}d^{\boldsymbol{A}\,}\cdot}=\nabla\cdot-\left[\mathbf{A},\cdot\right]$ is the covariant
differential (sometimes also denoted by $\nabla^{\mathbf{A}}$). Also, $A_0\in C^\infty(\Bbb{R}^3,\mathfrak{o})$ is the
$\mathfrak{o}$-valued scalar potential so that setting $\delta
H/\delta A_0\equiv 0$ produces the nonholonomic constraint
$\operatorname{div}^{\mathbf{A}} \mathbf{E}{+}\int \sigma f\,d^3 \mathbf{p}\, d\sigma=0$. The YM
field equations are
\begin{equation}
\partial_t\mathbf{E}{-}\operatorname{ad}^*_{A_0}\mathbf{E}+\int\sigma
f\,\left(\mathbf{p}-\langle\sigma,\mathbf{A}\rangle\right)d\sigma=\operatorname{curl}^\mathbf{A}\mathbf{B}
\,,\qquad\,
\partial_t \mathbf{A}={-}\mathbf{E}{-}d^\mathbf{A} A_0
\end{equation}
where $\mathbf{B}=d^\mathbf{A} \mathbf{A}$ is the magnetic component of the YM field and the
suffix $\bf A$ on differential operators always denotes covariant
differentiation \cite{GiHoKu1983}. In particular, one has $\operatorname{div}^\mathbf{A}\mathbf{E}=\operatorname{div}\mathbf{E}+\operatorname{ad}^*_{A_i}E_i$ and $\operatorname{curl}^\mathbf{A}\mathbf{B}=\operatorname{curl}\mathbf{B}-\epsilon_{ijk}\left[{A_i,B_k}\right]$, where summation over repeated indexes is intended.
 The above YM field equations are coupled to the
Vlasov equation
\begin{multline}
\frac{\partial f}{\partial
t}+\left(\mathbf{p}-\langle\sigma,\mathbf{A}\rangle\right)\cdot\frac{\partial
f}{\partial \mathbf{x}}-\left(\left\langle\sigma,\frac{\partial \mathbf{A}}{\partial
{\mathbf{x}}}\right\rangle\cdot\left(\mathbf{p}-\langle\sigma,\mathbf{A}\rangle\right)+\frac{\partial
A_0}{\partial \mathbf{x}}\right)\cdot\frac{\partial f}{\partial
\mathbf{p}}
\\
-\left\langle\sigma,\left[\left(\mathbf{p}-\langle\sigma,\mathbf{A}\rangle\right)\cdot
\mathbf{A}+A_0,\frac{\partial f}{\partial \sigma}\right]\right\rangle=0
\,,
\end{multline}
where we recognize that the Klimontovich particle solution \cite{Kl1982, GiHoKu1983}
\[
f(\mathbf{x,p},\sigma,t)=\delta(\mathbf{x}-\mathbf{Q}(t))\,\delta(\mathbf{p-P}(t))\,\delta(\sigma-\Upsilon(t))
\]
consistently returns Wong's equations for a single charge moving in
a Yang-Mills field \cite{Mo1984,MoMaRa84}.

 The equations of chromohydrodynamics \cite{GiHoKu1983} are
obtained from the equation hierarchy for the moments
\begin{equation}\label{def-CHD-moments}
X_{n,k}(\mathbf{x},t)=\int \mathbf{p}^n\,\sigma^k\,f(\mathbf{x,p},\sigma,t)\ {\rm d}p\,{\rm
d}\sigma
\end{equation}
by making use of the cold-plasma closure
\begin{equation}\label{COLDPLASMA}
f(\mathbf{x,p},\sigma,t)=X_{0,0}\,\delta\!\left(\mathbf{p}-\frac{\mathbf{X}_{1,0}}{X_{0,0}}\right)\,\delta\!\left(\sigma-\frac{X_{0,1}}{X_{0,0}}\right)
\end{equation}
Here, the zero-th moment $X_{0,0}$ is evidently the particle density, while $\mathbf{X}_{1,0}$ and $X_{0,1}$ are the fluid momentum and the Yang-Mills charge density respectively.
The moments $X_{n,k}$ possess a Poisson bracket \cite{GiHoTr2008}, which is obtained by {\color{black}using the chain rule}
\begin{equation}\label{chain-rule-moms}
\frac{\delta F}{\delta f}=\sum_{n,k}\frac{\delta X_{n,k}}{\delta f}\contract\frac{\delta F}{\delta X_{n,k}}=\sum_{n,k}\,p^n\,\sigma^k\contract\frac{\delta F}{\delta X_{n,k}}
\end{equation}
{(here $\contract$ denotes tensor contraction)} in the following Poisson structure
of the YM Vlasov equation \cite{GiHoKu1983}:
\begin{align}\label{YMVbraket}
\{F,K\}(f)=& \int f \left\{\frac{\delta F}{\delta f},\frac{\delta
K}{\delta f}\right\}{\rm d}^3\mathbf{x}\,{\rm d}p\,{\rm d}\sigma+\int
f\left\langle\sigma,\left[\frac{\partial}{\partial\sigma}\frac{\delta
F}{\delta f},\frac{\partial}{\partial\sigma}\frac{\delta K}{\delta
f}\right]\right\rangle{\rm d}^3\mathbf{x}\,{\rm d}p\,{\rm d}\sigma\,.
\end{align}
Then, retaining only the terms involving $D=X_{0,0}$, $G=X_{0,1}$
and $\mathbf{m=X}_{1,0}$ produces the following equations of
chromohydrodynamics \cite{GiHoKu1983}
\begin{align}\label{CHD}
\nonumber
\left(\frac{\partial }{\partial t}+\pounds_{\frac{\delta H}{\delta \mathbf{m}}}\right)D
&=0
\\
\nonumber \left(\frac{\partial }{\partial t}+\pounds_{\frac{\delta
H}{\delta \mathbf{m}}}\right)G &={-} \operatorname{ad}^*_{\frac{\delta
H}{\delta G}}G
\\
\left(\frac{\partial}{\partial t}+\pounds_{\frac{\delta H}{\delta
\mathbf{m}}}\right)\mathbf{m}&= -D\,\nabla\frac{\delta H}{\delta
D} - \left\langle G,\nabla\frac{\delta H}{\delta
G}\right\rangle
.
\end{align}
{\color{black}In this paper, Lie derivatives along
the vector field ${\delta H}/{\delta \mathbf{m}}$ are either denoted by $\pounds_{{\delta H}/{\delta \mathbf{m}}}$ or expressed explicitly, depending on convenience}. 
 The YM Vlasov Hamiltonian
transforms into
\begin{multline}
H=\frac12 \int \frac{1}{D}\left|\mathbf{m}-\left\langle
G,\mathbf{A}\right\rangle\right|^2d^3\mathbf{x}+\int D\,U(D)\,d^3\mathbf{x}
\\
-\int \left\langle G,A_0\right\rangle d^3\mathbf{x} +\frac12\int\left(
\left|d^\mathbf{A} \mathbf{A}\right|^2+\left|\mathbf{E}\right|^2\right) d^3\mathbf{x}-\int\left\langle
\mathbf{E},d^\mathbf{A}A_0\right\rangle d^3\mathbf{x}
\end{multline}
where the internal energy $D\,U(D)$ is inserted to account for pressure
effects. This yields the following equation for the velocity
$\mathbf{u}=D^{-1}\left(\mathbf{m}-\langle G, \mathbf{A}\rangle\right)$:
\begin{equation}\label{YM-velocity}
\frac{\partial \mathbf{u}}{\partial t}+\left(\mathbf{u}\cdot\nabla\right)\mathbf{u}=-D^{-1}\,\nabla
{\sf p}-D^{-1}\left\langle G,\mathbf{E}+\mathbf{u}\times \mathbf{B}\right\rangle
\end{equation}
where ${\sf p}=D^2\,dU/dD$. This equation is coupled to the Yang-Mills field equations
\begin{equation}\label{YM-fields}
\partial_t\mathbf{E}{-\operatorname{ad}^*_{A_0}\mathbf{E}+G\mathbf{u}}=\operatorname{curl}^\mathbf{A}\mathbf{B}
\,,\qquad\,
\partial_t \mathbf{A}={-}\mathbf{E}{-}d^\mathbf{A} A_0
\end{equation}
and the transport equations for the fluid density $D$ and the charge
density $G$:
\begin{equation}\label{YM-transport}
{\partial_t D}+\operatorname{div}\!\left(D\,\mathbf{u}\right)=0 \,,\qquad\
\partial_tG+\operatorname{ad}^*_{A_0}G+\operatorname{div}^{\bf A}\!\left(G\,\mathbf{u}\right)=0
\end{equation}
Here the ``hydrodynamic gauge'' $A_0+\mathbf{u\cdot A}=0$ is adopted. 

An important result concerning the CHD equations is that, in their
MHD limit \cite{HoKu1984}, their Hamiltonian structure is identical to the equations
for spin glasses, with the only difference residing in the explicit
expression of the Hamiltonian \cite{HoKu1988}. 
\rem{ %%%%%%%%%%%%%%%%%%%%%%%%%%%%%%%%%%%%%%%%%%%%%%
The MHD limit of CHD is
easily found \cite{HoKu1984} from two-species chromohydrodynamics, upon
imposing quasi-neutrality of the total YM charge and neglecting the
inertia of one species. As a result, the velocity equation of YM MHD
is \cite{HoKu1984}
\begin{equation}\label{YM-MHD-velocity}
{\color{black}D\left(\frac{\partial \mathbf{u}}{\partial t}+\left(\mathbf{u}\cdot\nabla\right)\mathbf{u}\right)=-\nabla
{\sf p}+\operatorname{curl}^\mathbf{A} \mathbf{B}\times \mathbf{B}}\,,
\end{equation}
while the other transport equations read
\begin{equation}\label{YM-MHD-others}
{\partial_t D}+\operatorname{div}\!\left(D\,\mathbf{u}\right)=0 \,;\qquad\,
{\partial_t G}+\operatorname{div}\!\left(G\,\mathbf{u}\right)= 0\,;\qquad\,
{\partial_t  \mathbf{A}}+\left(\mathbf{u}\cdot \nabla\right)\mathbf{A}+\nabla \mathbf{u}\cdot \mathbf{A}=0\,.
\end{equation}
Here the ``hydrodynamic gauge'' $A_0+\mathbf{u\cdot A}=0$ is adopted. 
The equivalence between the above equations and the
equations of spin glasses was established in \cite{HoKu1988}.
} %%%%%%%%%%%%%%%%%%%%%%%%%%%%%%%%%%%%%%%%%%%%%% 
{\color{black}Without entering the treatment of Yang-Mills MHD, the next section introduces the hybrid closure of the Yang-Mills-Vlasov distribution function. This closure is the basis for all the complex fluid models presented later.}

\subsection{Hybrid models}
This paper provides a general setting to include microscopic effects in the micromotion of complex fluids by the introduction of a new moment closure of the Yang-Mills Vlasov kinetic equation. {\color{black}This closure is} similar to the cold plasma closure \eqref{COLDPLASMA} that arises in the dynamics of quark-gluon plasmas in neutron stars. The motivation for this approach is the well known correspondence between many soft matter models and the fluid theory underlying Yang-Mills collisionless plasmas, {\color{black}which was reviewed in the previous section}. In particular the equations \eqref{CHD} of chromohydrodynamics  have been shown to apply also to complex fluids such as spin glasses and liquid crystals \cite{Ho2002,GayBRatiu}.

This paper develops  new theories of perfect complex fluids in which the transported order parameter field $\phi:\Bbb{R}^3\to M$ is replaced by  a \emph{position-dependent} probability distribution function ${\rho\in C^\infty(\Bbb{R}^3\times M)}$ on the order parameter space $M$. This statistical quantity arises very naturally in the context of kinetic equations of the type Vlasov-Yang-Mills, for which $M$ is a Lie group often called \emph{order parameter group} $\mathcal{O}$ \cite{Ho2002,GayBalmazTronci}. When particle dynamics is $\mathcal{O}$-invariant, simple Lie-Poisson reduction (i.e. $T^*\mathcal{O}/\mathcal{O}=\mathfrak{o}^*$) takes the dynamics on the cotangent bundle $T^*\mathcal{O}$ to the reduced dynamics on the dual Lie algebra $\mathfrak{o}^*$. Then, a new fluid model arises  from a moment closure of the Vlasov distribution function, {\color{black}which is defined} on $T^*(\Bbb{R}^3\times \mathcal{O})/\mathcal{O}\simeq \Bbb{R}^6\times\mathfrak{o}^*$ {(see previous section)}. In particular, if $f\in C^\infty(\Bbb{R}^6\times\mathfrak{o}^*)$ is the Vlasov distribution, one can introduce the cold-plasma-like closure
\begin{equation*}
f(\mathbf{x,p},\sigma,t)=\delta\!\left(\mathbf{p}-\frac{\int\! \mathbf{p} f\,\de\mathbf{p}\,\de\sigma}{\int\! f\,\de\mathbf{p}\,\de\sigma}\right)\,\int\!f\,\de\mathbf{p}\,.
\end{equation*}
One notices that the distribution $\rho(\mathbf{x},\sigma,t)=\int\!f\,\de\mathbf{p}$ belongs to a moment hierarchy that is different from the one { containing} the other two moments $\mathbf{m}(\mathbf{x},t)=\int\! \mathbf{p} f\,\de\mathbf{p}\,\de\sigma$ and $D(\mathbf{x},t)=\int\!  f\,\de\mathbf{p}\,\de\sigma$, since $\rho$ does not involve integration over $\sigma\in\mathfrak{o}^*$. %The fluid model generated by the above moment closure is called \emph{hybrid}, since it involves an ordinary fluid transporting a kinetic equation for $\rho$, which encodes the statistical information on the micromotion. Indeed,
From a mathematical point of view, the variable $\rho$ is a probability distribution function on $\mathfrak{o}^*$ that is attached to each fluid particle moving in $\Bbb{R}^3$. Hence, we write $\rho\in\mathrm{Den}(\Bbb{R}^3\times\mathfrak{o}^*)$, where $\mathrm{Den}$ denotes a space of {\color{black}probability densities}. Consequently, one can consider $\rho$ as a map belonging to the dual space of maps $\Bbb{R}^3\to C^\infty(\mathfrak{o}^*)$. On the other hand, $C^\infty(\mathfrak{o}^*)$ is identified with the Poisson algebra of $\mathcal{O}$-invariant Hamiltonian functions on $T^*\mathcal{O}$ (here { also} denoted by $C^\infty(T^*\mathcal{O})_\mathcal{O}$). Thus, we shall see that from a mathematical point of view, the introduction of the variable $\rho$ amounts to introducing the infinite-dimensional order parameter space $C^\infty(\mathfrak{o}^*)$, so that the geometric properties of complex fluid dynamics \cite{Ho2002,GayBRatiu} can be naturally extended to the case of infinite-dimensional order parameters.

The special feature of this approach is that the resulting fluid model naturally accounts for the statistical information on the micromotion. This is a statistical distribution which is dragged along the fluid particle trajectories, as it also happens for ordinary density variables in standard hydrodynamic models. Because of the coexistence of both the statistical distribution on $M$ and the ordinary fluid variables such as momentum and mass density, we call these models \emph{hybrid models}.
%\rem{ %%%%%%%%%%%%%%%%%%%%%%%%%
\noindent
\begin{figure}[h]
\center\it
\includegraphics[scale=.75]{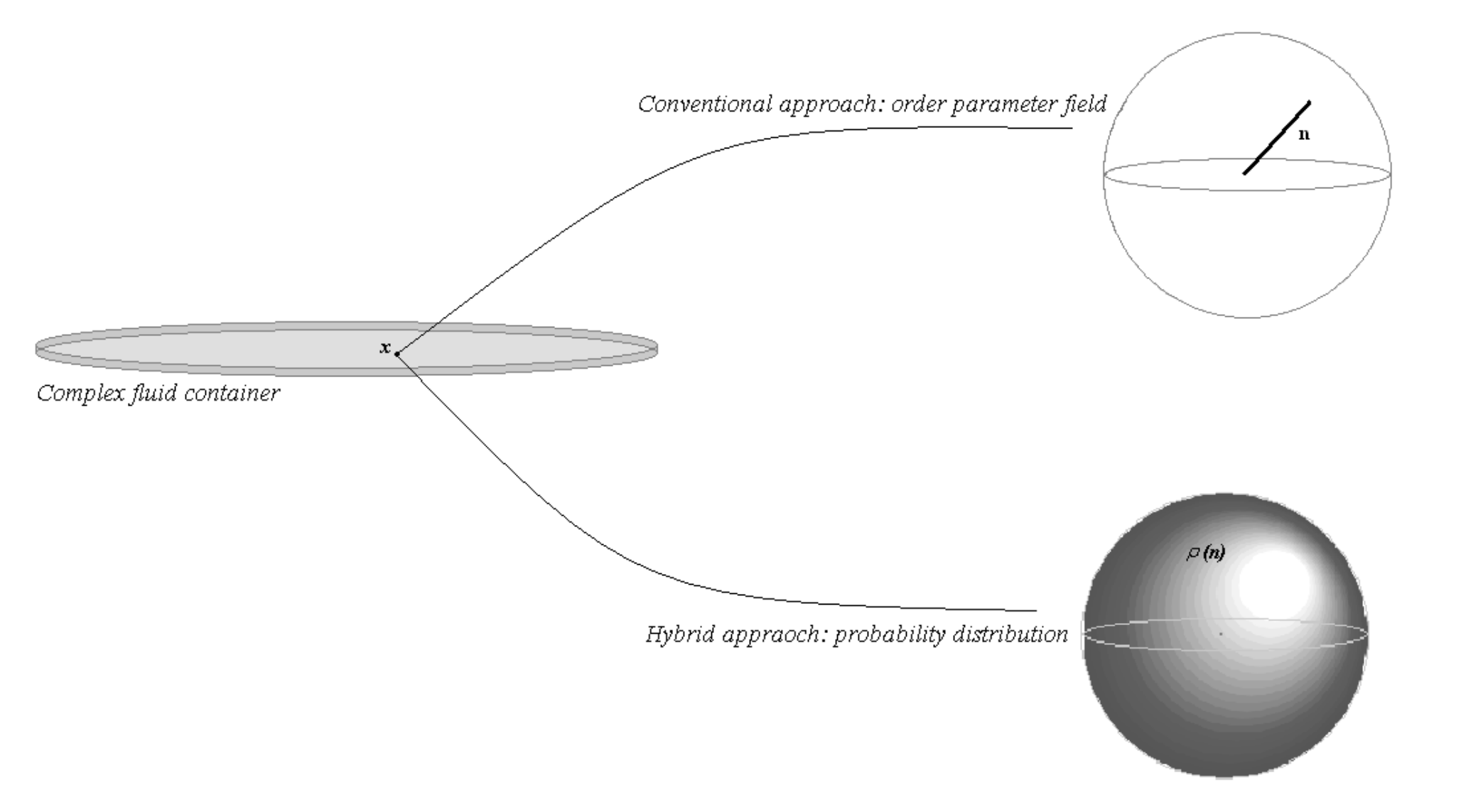}
\caption{Illustrative comparison of the conventional and the hybrid approaches with order parameter space $M=S^2$. While the conventional approach associates a director $\mathbf{n(x)}$ to each fluid parcel at position  $\mathbf{x}$, the hybrid approach associates the whole probability density $\rho(\mathbf{x},\boldsymbol{n})$. Probability densities of the same type also appear in Doi's theory of polymeric fluids \cite{DoEd1988} with negligible inertial effects.}
\end{figure}
%} %%%%%%%%%%%%%%%%%%%%%%%%%%%%

\paragraph{Plan of the paper.}
Section 3 presents the hybrid cold-plasma closure for Yang-Mills plasmas: the resulting Hamiltonian structure is explained in detail along with its momentum map property.  The whole theory is directly applied to magnetized fluids with multipolar order and the resulting dynamics is then compared to Isayev's equations \cite{Isayev}. Section 4 couples the hybrid closure of the Vlasov-Yang-Mills equation with the dynamics of the Yang-Mills field in order to produce the hybrid formulation of chromohydrodynamics. Then, the same hybrid closure is applied to the case of ferromagnetic hydrodynamics with quadrupolar interactions. Section 5 introduces the hybrid closure for systems with broken symmetry, such as nematic liquid crystals. In Section 6, the hybrid formulation of liquid crystal dynamics is presented, after deriving a suitable expression for the free energy which is inspired by the Frank energy. Then, the hybrid model of liquid crystals is shown to recover Eringen's formulation of micropolar liquid crystals \cite{Eringen}. Applications to cubatic liquid crystals are also proposed. This paper focuses on  Hamiltonian dynamics, thereby discarding the effects of dissipation. A geometric treatment of the dissipative dynamics in hybrid models is presented in \cite{HoPuTr09}.

\section{Hybrid closure of the Yang-Mills Vlasov equation}
One easily verifies that the hybrid cold-plasma closure of the YM Vlasov
equation
\begin{equation}
f(\mathbf{x,p},\sigma,t)=\rho(\mathbf{x},\sigma,t)\,\delta(\mathbf{p}-\mathbf{P(x},t))
\end{equation}
yields another Hamiltonian fluid system, which allows for a
microscopic statistical treatment of the YM charge dynamics. Indeed,
upon introducing the fluid momentum
\begin{equation}
\mathbf{m}(\mathbf{x},t)=\int \mathbf{p}\,f(\mathbf{x,p},\sigma,t)\ {\rm d}\mathbf{p}\,{\rm d}\sigma=\mathbf{P(x},t)\,\int \rho(\mathbf{x},\sigma,t)\ {\rm d}\sigma
\end{equation}
and by using the relation
\begin{equation}
\rho(\mathbf{x},\sigma,t)=\int f(\mathbf{x,p},\sigma,t)\ {\rm d}^3\mathbf{p}\,,
\end{equation}
direct substitution of the chain rule formula
\begin{equation}
\frac{\delta F}{\delta f}=\frac{\delta F}{\delta \rho}+\mathbf{p}\cdot\frac{\delta F}{\delta \mathbf{m}}
\end{equation}
into the YM Vlasov bracket {\eqref{YMVbraket}} yields the Poisson bracket for the
independent variables $(\rho,\mathbf{m})$
\begin{multline}
\{F,K\}(\mathbf{m},\rho)= \int \mathbf{m}\cdot \left[\frac{\delta F}{\delta
\mathbf{m}},\frac{\delta K}{\delta \mathbf{m}}\right]_\mathfrak{\!X}{\rm d}^3\mathbf{x}+\int
\rho\,\left\langle\sigma,\left[\frac{\partial}{\partial\sigma}\frac{\delta
F}{\delta \rho},\frac{\partial}{\partial\sigma}\frac{\delta
K}{\delta \rho}\right]\right\rangle{\rm d}^3\mathbf{x}\,{\rm d}\sigma
\\
- \int \rho\left(\pounds_{\delta F/\delta \mathbf{m}\,}\frac{\delta K}{\delta
\rho}-\pounds_{\delta K/\delta \mathbf{m}\,}\frac{\delta F}{\delta
\rho}\right){\rm d}^3\mathbf{x}\,{\rm d}\sigma\,, \label{PB-rho&m}
\end{multline}
where the Lie bracket $\left[\mathbf{V,V}'\right]_\mathfrak{X}:=(\mathbf{V}'\cdot\nabla)\mathbf{V}-(\mathbf{V}\cdot\nabla)\mathbf{V}'$ is minus the Jacobi-Lie bracket on the Lie algebra $\mathfrak{X}(\Bbb{R}^3)$ of vector fields on $\Bbb{R}^3$ \cite{MaRa,HoScSt}.
The above Poisson structure produces the equations
\begin{align}\nonumber
\left(\frac{\partial}{\partial t}+\pounds_{\delta H/\delta \mathbf{m}}\right)\mathbf{m}&=-\int\!\rho\,\nabla \frac{\delta H}{\delta \rho}\,\de\sigma
\\
\frac{\partial\rho}{\partial t}+{\rm div\!}\left(\rho\,\frac{\delta
H}{\delta \mathbf{m}}\right)&=-\left\langle\sigma,\left[\frac{\partial
\rho}{\partial \sigma},\frac{\partial}{\partial \sigma}\frac{\delta
H}{\delta \rho}\right]\right\rangle \,,
\label{newmodel}
\end{align}
which constitute the most fundamental hybrid micro-macroscopic model. It is
interesting to notice that the variables $(\rho,\mathbf{m})$ are moments of
the Vlasov distributions that belong to two different moment
hierarchies, thereby producing one more reason for the word
`hybrid'. This model provides a new philosophy of approaching order
parameter dynamics. Indeed, the macroscopic order parameter field is no more simply
transported along the fluid, but rather the fluid continuously
transports the whole statistical information about micromotion. This information
is entirely encoded in the probability distribution function $\rho$. This type of hybrid closure has recently been introduced in the context of Smoluchowski dynamics of ferromagnetic particles \cite{HoPuTr09}.
\rem{ %%%%%%%%%%%%%%%%%%%%%%%%%%%%%%%%%%
A possible explicit form of the Hamiltonian for the hybrid
equations, without involving the YM field equations, is given by
\begin{multline}
H=\frac12\int\frac{|m(\mathbf{x})|^2}{D(\mathbf{x})}+\frac12\int\!\rho(\mathbf{x},\sigma)\,\|\sigma\|^2\,d\sigma+\int D\,U(D)\,d\sigma\,d^3\mathbf{x}
\\
+\frac12\iint\!
\rho(\mathbf{x},\sigma)\,\Phi(\mathbf{x},q',\sigma,\sigma')\,\rho(\mathbf{x}',\sigma')\,d\sigma\,d\sigma'
\end{multline}
where $\|\cdot\|$ is a norm induced by a metric
$g\in\Omega^1(\mathfrak{o})\otimes\Omega^1(\mathfrak{o})$ on the Lie
algebra $\mathfrak{o}$, while the kernel $\Phi(\mathbf{x},\sigma)$
corresponds to some interaction potential and $D=\int\rho\,d\sigma$
is the particle density. Then, the hybrid closure yields the
following equations in terms of the velocity $v=m/D$:
\begin{align}\nonumber
\frac{\partial v}{\partial
t}+\left(v\cdot\nabla\right)v=&-\nabla{\sf
p}-\frac1{\int\rho(\mathbf{x},\sigma)\,d\sigma}\int\!\rho\,\nabla\Phi*\rho
\\
\frac{\partial\rho}{\partial t}+{\rm
div\!}\left(\rho\,v\right)=&-\left\langle\sigma,\left[\left(\Bbb{K}\sigma+\frac{\partial
\Phi}{\partial \sigma}*\rho\right),\frac{\partial \rho}{\partial
\sigma}\right]\right\rangle
\end{align}
where $\Phi*\rho$ denotes convolution, while the symmetric positive
matrix $\Bbb{K}$ is a map $\mathfrak{o}^*\to\mathfrak{o}$ arising
from the dual to the metric on $\mathfrak{o}$.
} %%%%%%%%%%%%%%%%%%%%%%%%%%%%%%%%%%

\begin{remark}[Lie-Poisson structure of the hybrid closure]\label{Rem-LPHybrid}
We notice that the Poisson bracket \eqref{PB-rho&m} is Lie Poisson on the dual of the semidirect-product Lie algebra
\[
\mathfrak{X}(\Bbb{R}^3)\,\circledS\,\mathcal{F}(\Bbb{R}^3,C^\infty(\mathfrak{o}^*))
\]
where $\mathcal{F}(\Bbb{R}^3,C^\infty(\mathfrak{o}^*))$ denotes the { space} of scalar functions taking values in the Poisson algebra  $C^\infty(\mathfrak{o}^*)$. Notice that this Lie algebra underlies the semidirect-product Lie group
\[
\operatorname{Diff}(\Bbb{R}^3)\,\circledS\,\mathcal{F}(\Bbb{R}^3,C^\infty(\mathfrak{o}^*))
\]
where the Lie algebra $\mathcal{F}(\Bbb{R}^3,C^\infty(\mathfrak{o}^*))$ also arises from an appropriate Lie group $\mathcal{F}(\Bbb{R}^3,{\mathcal{G}})$, where ${\mathcal{G}}\subset\operatorname{Diff}(T^*\mathcal{O})$ is a Lie group whose properties have been recently investigated in \cite{GBViTr}. The  Lie group ${\mathcal{G}}$ can be regarded as a subgroup of $\operatorname{Diff}_{\rm can}(T^*\mathcal{O})$, i.e. canonical transformations on $T^*\mathcal{O}$. {(More precisely, $\operatorname{Diff}_{\rm can}(T^*\mathcal{O})$ should be replaced here by the group of strict contact transformations on $T^*\mathcal{O}\times\Bbb{R}$, see \cite{GBViTr}).} This group $\mathcal{F}(\Bbb{R}^3,{\mathcal{G}})$ plays a similar role as the gauge group $\mathcal{F}(\Bbb{R}^3,\mathcal{O})$ in chromohydrodynamics \cite{GiHoKu1983}. See also the gauge theory treatment of complex fluids in \cite{DzVo}. However, in the present case ${\mathcal{G}}\subset\operatorname{Diff}_{\rm can}(T^*\mathcal{O})$ is infinite-dimensional, as opposed to $\mathcal{O}$, which is finite-dimensional.
\end{remark}

\subsection{Momentum map properties and singular solutions}

Since it is well known that moments of the Vlasov equation are momentum maps \cite{GiHoTr2008,HoTr2009}, it becomes natural to ask whether the moment quantity $\rho(\mathbf{x},\sigma)$ is a momentum map. In particular, we notice that the moment $\rho(f)$ is the zeroth-order moment of the moment hierarchy
\[
A_n(\mathbf{x},\sigma,t)=\int\mathbf{p}^{\otimes \,n}\,f\,\de^3\mathbf{p}\,,
\]
whose momentum map property arises as the dual of the Lie algebra homomorphism $\left\{\beta_n\right\}\mapsto\sum\mathbf{p}^{\otimes \,n\!}\contract\beta_n$, where $\contract$ denotes tensor contraction. This homomorphism associates to each symmetric contravariant tensor field $\left(\beta_n(\mathbf{x},\sigma)\right)^{i_1\dots i_n}\,\partial_{x^{i_1}}\dots\partial_{x^{i_n}}$ the Hamiltonian function $h(\mathbf{x,p},\sigma)=\sum\mathbf{p}^{\otimes \,n\!}\contract\beta_n$. Setting $n=0$ determines the Lie algebra inclusion $i:\mathcal{F}(\Bbb{R}^3,C^\infty(\mathfrak{o}^*))\hookrightarrow C^\infty(T^*\Bbb{R}^3\times\mathfrak{o}^*)$, such that $i(\beta_0)=\beta_0$. Since the latter is evidently a Lie algebra homomorphism, then the dual map $i^*(f)=\int\! f\,\de^3\mathbf{p}=A_0(\mathbf{x},\sigma)$ is an equivariant momentum map.

\rem{ %%%%%%%%%%%%%%%%%%%%%%%%%%%%%%%%%
This question can be partially answered by using the definition of momentum map \cite{HoScSt,MaRa}. Thus, one computes
\begin{multline*}
\int \!f \left\{\frac{\delta F}{\delta f},\frac{\delta}{\delta f}\!\int\! \rho(f)\,\xi(\mathbf{x},\sigma)\,\de^3\mathbf{x}\,\de^3\sigma\right\}{\rm d}^3\mathbf{x}\,{\rm d}p\,{\rm d}\sigma+\int\!
f\left\langle\sigma,\left[\frac{\partial}{\partial\sigma}\frac{\delta
F}{\delta f},\frac{\partial}{\partial\sigma}\frac{\delta}{\delta f}\!\int\! \rho(f)\,\xi(\mathbf{x},\sigma)\,\de^3\mathbf{x}\,\de^3\sigma\right]\right\rangle{\rm d}^3\mathbf{x}\,{\rm d}p\,{\rm d}\sigma
\\
=
\int f \left\{\frac{\delta F}{\delta f},\xi(\mathbf{x},\sigma)\right\}{\rm d}^3\mathbf{x}\,{\rm d}p\,{\rm d}\sigma+\int
f\left\langle\sigma,\left[\frac{\partial}{\partial\sigma}\frac{\delta
F}{\delta f},\frac{\partial}{\partial\sigma}\xi(\mathbf{x},\sigma)\right]\right\rangle{\rm d}^3\mathbf{x}\,{\rm d}p\,{\rm d}\sigma
\\
=
\int \frac{\delta F}{\delta f} \left(\left\{\xi(\mathbf{x},\sigma),f\right\}+\left\langle\operatorname{ad}^*_{\partial\xi/\partial\sigma}\sigma,\frac{\partial f}{\partial\sigma}\right\rangle\right){\rm d}^3\mathbf{x}\,{\rm d}p\,{\rm d}\sigma
\end{multline*}
}    %%%%%%%%%%%%%%%%%%%%%%%%%%%%%%%%%

In order to understand the physical meaning of the hybrid closure,
it is helpful to notice some relevant solutions of the kinetic equation for $\rho$ in \eqref{newmodel}. First, one simply verifies that the solution
\[
\rho(\mathbf{x},\sigma,t)=D(\mathbf{x},t)\ \delta\!\left(\sigma-\frac{G(\mathbf{x},t)}{D(\mathbf{x},t)}\right)
\]
yields the CHD equation for the macroscopic YM charge density $G\in C^\infty(\Bbb{R}^3,\mathfrak{o})^*$, while the Klimontovich single-particle solution
\begin{equation}\label{klim}
\rho(\mathbf{x},\sigma,t)=\delta\!\left(\mathbf{x-Q}(t)\right)\,\delta\!\left(\sigma-\Upsilon(t)\right)
\end{equation}
yields
\[
\dot{\mathbf{Q}}= \left.\frac{\delta H}{\delta \mathbf{m}}\right\vert_{(\mathbf{x},\sigma)=(\mathbf{x}(t),\Upsilon(t))}
\,,\qquad\
\dot{\Upsilon}=\left.\operatorname{ad}^*_{\,{\delta H}/{\delta \rho}}\Upsilon\right\vert_{(\mathbf{x},\sigma)=(\mathbf{x}(t),\Upsilon(t))}
\]
thereby returning the fluid particle trajectory $\mathbf{Q}(t)$ together with the micromotion for its YM charge $\Upsilon(t)$.
Another relevant class of singular solutions is provided by the expression
\begin{equation}\label{newsingsol}
\rho(\mathbf{x},\sigma,t)=\int \!w(\sigma,s,t)\,\delta(\mathbf{x-Q}(s,t))\,d^k s\,.
\end{equation}
Here  $\mathbf{Q}:S\hookrightarrow\Bbb{R}^3$  is a time dependent map 
(embedding, $\mathbf{Q}\in\operatorname{Emb}(S,\Bbb{R}^3)$) from a lower dimensional submanifold
$S\subseteq\Bbb{R}^3$ into the physical space $\Bbb{R}^3$. On the
other hand, the quantity $w$ is interpreted as a map $w:S\to{
\mathrm{Den}(\mathfrak{o}^*)}$ from $S$ into the infinite-dimensional
space ${
\mathrm{Den}(\mathfrak{o}^*)}$ of probability densities on the
dual Lie algebra $\mathfrak{o}^*$. Direct substitution of the above
solution ansatz yields the following equations
\begin{align}\nonumber
\frac{\partial \mathbf{Q}(s,t)}{\partial t}=& \left.\frac{\delta H}{\delta \mathbf{m}}\right\vert_{\mathbf{{
x}=Q}(s,t)}
\\
\frac{\partial w(\sigma,s,t)}{\partial t}=&-\left\langle\sigma,\left[\frac{\partial w(\sigma,s,t)}{\partial \sigma},\left.\frac{\partial}{\partial \sigma}\frac{\delta H}{\delta \rho}\right|_{\mathbf{x=Q}(s,t)}
\right]\right\rangle
\end{align}
For example, if $\operatorname{dim}S=1$, the above solutions describe the evolution of a charged filament which supports the statistical distribution $w(\sigma,s,t)$ for the dynamics of the YM charge along the filament itself. In the simplest case when $\operatorname{dim}S=0$, the label $s$ is lost in \eqref{newsingsol} and this solution reduces to a point particle with position $\mathbf{Q}(t)$, supporting a distribution $w(\sigma,t)$ for the dynamics of its own YM charge. The latter example yields a system of equations composed of the ODE for the position coordinate $\mathbf{Q}$ and the PDE for the probability density $w$ and this is a unique feature of the hybrid closure, whose singular solutions may be  different from the usual Klimontovich particle solution \eqref{klim}.

\begin{remark}[{\color{black}Singular solutions are momentum maps}]
Notice that the singular solution \eqref{newsingsol} arises as a momentum map
\[
T^{*\!}\operatorname{Emb}(S,\Bbb{R}^3)\times\mathcal{F}(S,C^\infty(\mathfrak{o}^*))^*\to\mathcal{F}(\Bbb{R}^3,C^\infty(\mathfrak{o}^*))^*
\]
where the Poisson manifold $P=T^{*\!}\operatorname{Emb}(S,\Bbb{R}^3)\times\mathcal{F}(S,C^\infty(\mathfrak{o}^*))^*$ is endowed with the Poisson bracket
\begin{equation}
\left\{F,K\right\}=\int\!\left(\frac{\delta F}{\delta\mathbf{Q}}\cdot\frac{\delta K}{\delta\mathbf{P}}-\frac{\delta K}{\delta\mathbf{Q}}\cdot\frac{\delta F}{\delta\mathbf{P}}\right)\de s+\int w\left\langle\sigma,\left[\frac{\partial}{\partial\sigma}\frac{\delta F}{\delta w},\frac{\partial}{\partial\sigma}\frac{\delta K}{\delta w}\right]\right\rangle\de s\,\de\sigma\,.
\end{equation}
The infinitesimal generator of the associated action reads as
\[
\xi_P[F]=-\int\!\frac{\delta F}{\delta\mathbf{P}}\cdot\frac{\partial}{\partial\mathbf{Q}}\int w(s,\sigma)\,\xi(\mathbf{Q}(s),\sigma)\,\de\sigma+\int \frac{\delta F}{\delta w}\left\langle\operatorname{ad}^*_{\,\partial \xi(\mathbf{Q},\sigma)/\partial\sigma}\sigma,\frac{\partial w}{\partial\sigma}\right\rangle\de s\,\de\sigma\,,
\]
which arises from the group action
\begin{align*}
\Big({\bf Q}^{\,(t)} ,\,{\bf P}^{\,(t)},\, w^{\,(t)}\Big)&=
\left({\bf Q}^{\,(0)} ,\, {\bf P}^{\,(0)}+\,\left(\big\langle w^{(0)},\left(\nabla\!\exp(-t\xi)\right)\exp(t\xi)\big\rangle\circ{\bf Q}^{\,(0)}\right)
,\, {\sf Ad}^*_{\,\exp\left(\,t\,\xi\right)}\ w^{(0)}
\right)
\,,
\end{align*}
where ${\sf Ad^*}$ is the group coadjoint operator on the Lie group ${\mathcal{G}}$ underlying the Lie algebra $C^\infty(\mathfrak{o}^*)$, such that $C^\infty(\mathfrak{o}^*)=T_e{\mathcal{G}}$. Similarly, the notation $\left(\nabla g^{-1}\right)g$ makes use of the tangent lift of the right multiplication in ${\mathcal{G}}$. See \cite{GBViTr} for more details on how the group ${\mathcal{G}}$ is constructed. It is interesting to notice that the above construction coincides with the momentum map construction in \cite{HoTr2008} (in the absence of fluid motion) for geodesic flows on semidirect-product Lie groups, with the only difference that the group ${\mathcal{G}}$ is infinite-dimensional in the present case.
\end{remark}

\subsection{The mass density}\label{Hybrid+Density}
Although the hybrid equations \eqref{newmodel} are {\color{black}physically consistent}, in practical situations it is convenient to allow for the mass density
\begin{equation}\label{zerolevelset}
D(\mathbf{x},t)=\int \rho(\mathbf{x},\sigma,t)\ {\rm d}\sigma
\end{equation}
to be a dynamical variable. This is possible upon writing the cold-plasma solution as
\begin{equation}\label{Tenerife}
f(\mathbf{x,p},\sigma,t)=\rho(\mathbf{x},\sigma,t)\,\delta\!\left(\mathbf{p}-\frac{\mathbf{m}(\mathbf{x},t)}{D(\mathbf{x},t)}\right)
,
\end{equation}
so that the variables $(D,\rho,\mathbf{m})$ are treated as three different variables.
In this case, substitution of the chain rule formula
\begin{equation}
\frac{\delta F}{\delta f}=\frac{\delta F}{\delta D}+\frac{\delta F}{\delta \rho}+\mathbf{p}\cdot\frac{\delta F}{\delta \mathbf{m}}
\end{equation}
in \eqref{YMVbraket} yields the Poisson bracket
\begin{align}\nonumber
\{F,K\}(\mathbf{m},\rho)=& \int \mathbf{m}\cdot
\left[\frac{\delta F}{\delta \mathbf{m}},\frac{\delta K}{\delta
\mathbf{m}}\right]_\mathfrak{\!X}{\rm d}^3\mathbf{x}+\int \rho\,\left\langle\sigma,\left[\frac{\partial}{\partial\sigma}\frac{\delta
F}{\delta \rho},\frac{\partial}{\partial\sigma}\frac{\delta K}{\delta \rho}\right]\right\rangle{\rm d}^3\mathbf{x}\,{\rm d}\sigma
\\
\nonumber
&- \int \rho\left(\pounds_{\delta F/\delta m\,}\frac{\delta
K}{\delta \rho}-\pounds_{\delta K/\delta m\,}\frac{\delta F}{\delta
\rho}\right){\rm d}^3\mathbf{x}\,{\rm d}\sigma
\\
&- \int D\left(\pounds_{\delta F/\delta m\,}\frac{\delta
K}{\delta D}-\pounds_{\delta K/\delta m\,}\frac{\delta F}{\delta
D}\right){\rm d}^3\mathbf{x}\,,
\label{PBwithD}
\end{align}
The resulting equations of motion read as
\begin{align}\label{HybEq+D1}
\left(\frac{\partial}{\partial t}+\pounds_{\delta H/\delta \mathbf{m}}\right)\mathbf{m}&=-D\nabla\frac{\delta H}{\delta D}-\int\!\rho\,\nabla \frac{\delta H}{\delta \rho}\,\de\sigma
\\
\nonumber
\frac{\partial\rho}{\partial t}+{\rm div\!}\left(\rho\,\frac{\delta H}{\delta \mathbf{m}}\right)&=-\left\langle\sigma,\left[\frac{\partial \rho}{\partial \sigma},\frac{\partial}{\partial \sigma}\frac{\delta H}{\delta \rho}\right]\right\rangle
\\
\frac{\partial D}{\partial t}+{\rm div\!}\left(D\,\frac{\delta H}{\delta \mathbf{m}}\right)&=0\,,
\label{HybEq+D3}
\end{align}
which are accompanied by the constraint \eqref{zerolevelset}. The latter constraint belongs to a geometric class of constraints arising as the zero-level set of a certain momentum map \cite{MaRa,HoScSt}. Gauss' law in electromagnetism is a celebrated example {\color{black}of this construction} \cite{MaWeRaScSp}.

It is easy to derive the Kelvin circulation dynamics associated to equation \eqref{HybEq+D1}. To see this, we begin by considering the general relation \cite{MaRa,HoScSt}
\[
\frac{d}{dt}\!\left(\mathrm{Ad}^*_\mathsf{g\,}\mu\right)=\mathrm{Ad}^*_\mathsf{g}\!\left(\frac{d\mu}{dt}+\mathrm{ad}^*_{\dot{\mathsf{g}}\mathsf{g}^{-1}\,}\mu\right),
\]
which holds for any Lie group element $\mathsf{g}\in{\mathscr{G}}$ and any dual vector $\mu\in\mathfrak{g}^*$ in the dual of the Lie algebra $\mathfrak{g}=T_e{\mathscr{G}}$. Then, by the Lie-Poisson nature of the bracket \eqref{PBwithD}, one can write equation \eqref{HybEq+D1} in the form 
\[
\frac{d}{dt}\left(\eta^*\mathbf{m}\right)=\eta^*\!\left(\frac{\partial\mathbf{m}}{\partial t}+\pounds_{\left(\dot{\eta}\circ\eta^{-1}\right)\,}\mathbf{m}\right)
=
-\eta^*\!\left(D\nabla\frac{\delta H}{\delta D}+\int\!\rho\,\nabla \frac{\delta H}{\delta \rho}\,\de\sigma\right)
\]
where the star symbol denotes pullback and $\eta\in\mathrm{Diff}(\Bbb{R}^3)$ is a diffeomorphism such that $\delta H/\delta\mathbf{m}=\dot\eta\circ\eta^{-1}$. Upon dividing by $D_0=\eta^*D$, taking the circulation over a closed loop ${\mathcal{C}}_0$ yields
\[
\oint_{{\mathcal{C}}_0}\frac1{D_0}\frac{d}{dt}\,\eta^*\!\left(\mathbf{m}\cdot\de\mathbf{x}\right)=-\oint_{{\mathcal{C}}_0}\frac1{D_0}\,\eta^*\!\left(\int\!\rho\,\nabla\frac{\delta H}{\delta\rho}\,\de\sigma\cdot\de\mathbf{x}\right)
\]
Thus, changing variables produces the following circulation dynamics:
\[
\frac{d}{dt}\oint_{{\mathcal{C}}}\frac{\mathbf{m}}D\cdot\de\mathbf{x}=-\oint_{{\mathcal{C}}}\frac1D\!\left(\int\!\rho\,\nabla\frac{\delta H}{\delta\rho}\,\de\sigma\!\right)\cdot\de\mathbf{x}\,,
\]
where ${\mathcal{C}}=\eta\circ{\mathcal{C}}_0$ is a loop moving with the fluid velocity $\mathbf{u}=\dot\eta\circ\eta^{-1}=\delta H/\delta\mathbf{m}$.
 When the Hamiltonian $H$ depends only on the first-order moment $G=\int\!\sigma\rho\,\de\sigma$, then the above relation recovers the circulation dynamics previously found in \cite{Ho2002} for the Lagrangian dynamics of perfect complex fluids.
\rem{ %%%%%%%%%%%%%%%%%%%%%%%%%%%%%%%%%%%%%%%%%%%%%
 On the other hand, when higher order moments $A_n=\int\!\sigma^{\otimes n}\rho\,\de\sigma$ are considered, the above Kelvin circulation dynamics becomes
\[
\frac{d}{dt}\oint_{\gamma}\frac{\mathbf{m}}D\cdot\de\mathbf{x}=-\oint_{\gamma}\frac1D\sum_{n>0}\!\left(\int\!A_n\contract\,\nabla\frac{\delta H}{\delta A_n}\,\de\sigma\!\right)\cdot\de\mathbf{x}\,,
\]
where the contraction symbol $\contract$ is such that all upper and lower tensor indexes are contracted, that is $A_n\contract\nabla\beta_n=(A_n)_{\nu_1\dots \nu_n}\nabla(\beta_n)^{\nu_1\dots\nu_n}$.
}    %%%%%%%%%%%%%%%%%%%%%%%%%%%%%%%%%%%%%%%%%%%%%

Moreover, notice that the equations \eqref{HybEq+D1}-\eqref{HybEq+D3} leave the following quantity invariant:
\[
C(D,\rho)=\int \!D\,\Lambda(D^{-1}\rho)\,\de\sigma\,\de \mathbf{x}\,,
\]
for any scalar function $\Lambda$.
More particularly, the above quantity is a Casimir invariant for the Poisson bracket \eqref{PBwithD}. This is easily seen by evaluating the bracket
\begin{align*}
\{F,C\}=&\int \!\rho\left\langle\sigma,\left[\frac{\partial}{\partial\sigma}\frac{\delta
F}{\delta \rho},\frac{\partial\Lambda'}{\partial\sigma}\right]\right\rangle{\rm d}^3\mathbf{x}\,{\rm d}\sigma
- \int \!\rho\, \pounds_{\delta F/\delta \mathbf{m}}\Lambda'{\rm d}^3\mathbf{x}\,{\rm d}\sigma
- \int \!D\, \pounds_{\delta F/\delta \mathbf{m}\!}\int\!\left(\Lambda -\rho D^{-1\!}\Lambda'\right){\rm d}\sigma\,{\rm d}^3\mathbf{x}
\\
=&\int \!\Lambda'\left\langle\sigma,\left[\frac{\partial\rho}{\partial\sigma},\frac{\partial}{\partial\sigma}\frac{\delta
F}{\delta \rho}\right]\right\rangle{\rm d}^3\mathbf{x}\,{\rm d}\sigma
+ \int \!\Lambda'\Big( \!\pounds_{\delta F/\delta \mathbf{m}\,}\rho-D\, \pounds_{\delta F/\delta \mathbf{m}\,}\left(\rho D^{-1}\right)
- \rho D^{-1} \pounds_{\delta F/\delta \mathbf{m}\,} D\Big) \,{\rm d}^3\mathbf{x}\,{\rm d}\sigma
\\
=&\int \!D\left\langle\sigma,\left[D^{-1}\Lambda'\,\frac{\partial\rho}{\partial\sigma},\frac{\partial}{\partial\sigma}\frac{\delta
F}{\delta \rho}\right]\right\rangle{\rm d}^3\mathbf{x}\,{\rm d}\sigma
+ \int \!\Lambda'\Big( \!\pounds_{\delta F/\delta \mathbf{m}\,}\rho- \pounds_{\delta F/\delta \mathbf{m}\,}\left(\rho\, D\,D^{-1}\right)\Big)
{\rm d}^3\mathbf{x}\,{\rm d}\sigma
\\
=&\int \!\left\langle\sigma,\left[\frac{\partial D\Lambda}{\partial\sigma},\frac{\partial}{\partial\sigma}\frac{\delta
F}{\delta \rho}\right]\right\rangle{\rm d}^3\mathbf{x}\,{\rm d}\sigma
+ \int \!\Lambda'\left( \!\pounds_{\delta F/\delta \mathbf{m}\,}\rho- \pounds_{\delta F/\delta \mathbf{m}\,}\rho\right)
{\rm d}^3\mathbf{x}\,{\rm d}\sigma
\\
=&0\,,
\end{align*}
which holds for \emph{any} functional $F(\mathbf{m},D,\rho)$. {\color{black}The prime notation above $\Lambda'$ denotes total derivative of $\Lambda$ with respect to its argument.} Thus, the hybrid closure possesses a large family of Casimir functionals {\color{black}that can be used for nonlinear stability studies \cite{HoMaRaWe1985}.}

{
\begin{remark}[Alternative form of the hybrid closure]

The previous section showed how the mass density $D$ can be treated as an extra dynamical variable. However, this is possible only because of the constraint \eqref{zerolevelset}. This constraint prevents the variables $D$ and $\rho$ from being mutually independent, which instead would be preferable in a dynamical model. This problem can be easily avoided upon introducing the variable
\[
\varphi(\mathbf{x},\sigma,t)=\frac{\rho(\mathbf{x},\sigma,t)}{D(\mathbf{x},t)}
\]
which is a \emph{scalar function} on physical space $\Bbb{R}^3$ taking values in the \emph{distributions on the dual Lie algebra} $\mathfrak{o}^*$. 
\rem{ %%%%%%%%%%%%%%%%%%%%%%%%%%%%%%%%
\begin{remark}[Infinite-dimensional order parameter space]
The variable $\varphi$ can be considered as a time-dependent map
\[
\varphi:\Bbb{R}^3\to\operatorname{Den}(\mathfrak{o}^*)\,,
\]
where $\operatorname{Den}(\mathfrak{o}^*)$ denotes distributions on $\mathfrak{o}^*$. In this situation, the space $\operatorname{Den}(\mathfrak{o}^*)$ is interpreted as an \emph{infinite-dimensional order parameter space}. This case differs from all common situations encountered in condensed matter physics, where the order parameter space is given by a finite-dimensional coset space. Also, notice that the space $\operatorname{Den}(\mathfrak{o}^*)$ is dual to the Lie algebra of scalar functions $C^\infty(\mathfrak{o}^*)$. This Lie algebra underlies a rather complicated Lie group, which has been recently identified in \cite{GBViTr}.
\end{remark}
}  %%%%%%%%%%%%%%%%%%%%%%%%%%%%%%%%
Compatibility with the constraint \eqref{zerolevelset} requires that these distributions are normalized to unity at each point, that is
\[
\int \!\varphi(\mathbf{x},\sigma,t)\,\de\sigma=1\,,\quad\forall \,\mathbf{x}\in\Bbb{R}^3,\ \forall \,t\in\Bbb{R}\,.
\]
This is easily interpreted by saying that each fluid particle carries a probability distribution on $\mathfrak{o}^*$ such that the total probability is the always the same for all particles.
%%%%%%%%%%%%%%%%%%%%%%%%%%%%%%%%%%%%%%%%%%%%%%%%%%%%%%%%%%%%%%%%%%
All the previous considerations concerning the equations \eqref{newmodel} also apply to the above dynamics without essential modifications. In particular, the existence of singular solutions is preserved by this change of variables.
\end{remark}
}

\subsection{Hybrid hydrodynamics of spin nematic phases}
{ While} the magnetization dynamics has been addressed a great attention in various contexts, from ferrofluids to spin glasses, the emergence of other magnetic phases with multipolar order has posed relevant problems towards dynamical modeling. {\color{black}For example, magnets with multipolar interactions have been recently studied in \cite{SuLuLa}, where different multipolar ferromagnetic phases were presented together with their phase diagram. In particular, magnets with quadrupolar order are known under the name of \emph{spin nematic phases} \cite{PeLa2011}, since their order parameter is similar in form to that adopted for nematic liquid crystals.}

An interesting {\color{black}attempt for the introduction of quadrupole order in the dynamics} was carried out in the late 90's by Isayev and collaborators \cite{Isayev,IsKoPe,DeKo}, who developed a Poisson bracket approach to include the traceless symmetric tensor
\[
Q=\left\langle\boldsymbol\sigma\otimes\boldsymbol\sigma-\frac23|\boldsymbol\sigma|^2\mathbf{I}\right\rangle
\,,
\]
as an extra dynamical variable. Here, $\mathbf{I}$ is the identity matrix and the angle brackets $\left\langle\cdot\right\rangle$ denote an averaging process over the space of spin, so that $\boldsymbol\sigma\in\Bbb{R}^3$ is the microscopic spin variable. {(No confusion with the pairing bracket $\left\langle\cdot,\cdot\right\rangle$ should arise). Strictly speaking, one has $\boldsymbol\sigma\in\mathfrak{su}(2)\simeq\mathfrak{so}(3)$; however, in this paper we shall use the hat map $\widehat{\sigma}_{ij}=-\epsilon_{ijk}\sigma_k$ to identify $\mathfrak{so}(3)\simeq\Bbb{R}^3$. Isayev's theory enlarges the spin symmetry to consider coadjoint motion on the Lie group $SU(3)$ of special unitary transformations.  As we shall see, this approach does not capture essential dynamical features, since it returns trivial dynamics for $Q$. Also,} this approach is not feasible when higher order interactions must be considered and the use of higher-order tensors becomes necessary. For example, higher order multipolar interactions have been recently studied in \cite{SuLuLa}, where different multipolar ferromagnetic phases were presented together with their phase diagram.

From the preceding sections, it is clear that multipolar correlations are naturally encoded by hybrid models, which replace the magnetization order parameter by the distribution on the space of spin. For example, in the simplest case when both disclinations and external magnetic fields are absent, one can apply the Poisson bracket \eqref{PBwithD} to the following Hamiltonian function
\begin{equation}\label{Magnet-Hamiltonian}
{\color{black}H(\mathbf{m},\rho)=\frac12\int\! |\mathbf{u}|^2\,\de^3\mathbf{x}\,\de\sigma+\frac1{2\chi}\int\left|\boldsymbol{\mathcal{M}}(\rho)\right|^{2}\de^3\mathbf{x}
+\frac\mu2\!\int\!\operatorname{Tr}\!\left(Q^2(\rho)\right)\de^3\mathbf{x}
%+
%\lambda\!\int\!\operatorname{Tr}\!\left(Q(\rho)\boldsymbol{\mathcal{M}}(\rho)\boldsymbol{\mathcal{M}}^T(\rho)\right)\de^3\mathbf{x}
}
\,,
\end{equation}
which is the natural extension of the { spin nematic} Hamiltonian in \cite{AnGr} to the case of an incompressible fluid with magnetization $\boldsymbol{\mathcal{M}}=\int\!\rho\,\boldsymbol\sigma\,\de\sigma$. { Here, $\chi$ and $\mu$ are left as unspecified physical constants.} Here, the average defining the tensor $Q$ is calculated with respect to the distribution $\rho$, so that
\begin{equation}\label{Q-spintensor}
Q(\rho)=\int\!\rho(\boldsymbol\sigma)\left(\boldsymbol\sigma\otimes\boldsymbol\sigma-\frac23|\boldsymbol\sigma|^2\mathbf{I}\right)\de\sigma\,.
\end{equation}
{\color{black}The Hamiltonian \eqref{Magnet-Hamiltonian}} can be regarded as the total energy for a (incompressible) {\color{black}spin nematic liquid without disclinations. Notice that inserting \eqref{Magnet-Hamiltonian} in the first equation of \eqref{newmodel} yields Euler's fluid equation
\[
{
\frac{\partial\mathbf{u}}{\partial t}+(\mathbf{u}\cdot\nabla)\mathbf{u}=-\nabla\mathsf{p}
}
\]
 for the divergence-less velocity $\mathbf{u}$ and this  is due to the absence of both defects and external fields (which would produce non-vanishing force terms). On the other hand, upon using the Hamiltonian \eqref{Magnet-Hamiltonian} in the second equation of \eqref{newmodel}, one obtains the following kinetic equation on the spin space}
\begin{equation}\label{Rho-magnets}
\frac{\partial \rho}{\partial t}+\mathbf{u}\cdot\nabla\rho=g\,\frac{\partial \rho}{\partial \boldsymbol\sigma}\cdot\boldsymbol\sigma\times\left(\chi^{-1}\boldsymbol{\mathcal{M}}+2\mu Q\boldsymbol\sigma\right),
\end{equation}
where the gyromagnetic factor $g$ has been introduced.
{\color{black}Notice that standard tensor-index computations yield the pure transport relation $\partial_{t}\boldsymbol{\mathcal{M}}+(\mathbf{u}\cdot\nabla)\boldsymbol{\mathcal{M}}=0$, so that the initial condition $\boldsymbol{\mathcal{M}}_0=0$ reproduces  spin nematic behavior \cite{AnGr,LaDoLhSiTr}. The same magnetization dynamics also arises from Isayev's theory \cite{Isayev}, upon using the total energy \eqref{Magnet-Hamiltonian}. Moreover, upon taking the second-order moment of \eqref{Rho-magnets}, one obtains}
\begin{equation}\label{Q-equation}
\frac{\partial {Q}_{\alpha\beta}}{\partial t}+\mathbf{u}\cdot\nabla{Q}_{\alpha\beta}=\frac{g}\chi\left[\widehat{\mathcal{M}}\,,{Q}\right]_{\alpha\beta}-2g\mu\,{Q}_{\gamma\nu}\big(\left\langle\sigma_\beta\sigma_\nu\sigma_\mu\right\rangle\epsilon_{\alpha\mu\gamma}+\left\langle\sigma_\alpha\sigma_\nu\sigma_\mu\right\rangle\epsilon_{\beta\mu\gamma}\big)\,.
\end{equation}
{\color{black}where we have introduced the hat map notation $\widehat{\mathcal{M}}_{ij}:=-\epsilon_{ijk}\mathcal{M}_k$.}

Although the matrix commutator on the right hand side of \eqref{Q-equation} also emerges in Isayev's corresponding equation \cite{Isayev}
\begin{equation}\label{Q-equation-Isayev}
\frac{\partial {Q}^{(I)}_{\alpha\beta}}{\partial t}+\mathbf{u}\cdot\nabla{Q}^{(I)}_{\alpha\beta}=\frac{g}\chi\left[\widehat{\mathcal{M}}\,,{Q}^{(I)}\right]_{\alpha\beta}+g\frac{\mu}2\,{Q}^{(I)}_{\gamma\nu}\big(\delta_{\gamma\beta}\epsilon_{\nu\mu\alpha}+\delta_{\alpha\nu}\epsilon_{\gamma\beta\mu}\big)\mathcal{M}_{\mu}
\,,
\end{equation}
the latter replaces the other terms by a combination of the second and first moments, { which leads to the trivial dynamics $\partial_t{Q}^{(I)}+\mathbf{u}\cdot\nabla{Q}^{(I)}=0$ for spin nematics with vanishing  magnetization. On the other hand, equation \eqref{Q-equation} requires computing the evolution of the moment hierarchy and special closure procedures need to be invoked. The study of these closure procedures and their application to realistic magnetic media will be the subject of future work. Another immediate advantage of the present hybrid setting is that it easily allows to incorporate the effects of higher order multipolar interactions (like those recently studied in \cite{SuLuLa}), whose corresponding terms would appear in the expression of the Hamiltonian.  Notice that the family of Casimirs $C=\int\!\Lambda(\rho)\,\de\mathbf{x}\,\de\sigma$ allows a nonlinear stability analysis, which will be treated elsewhere.
}
%The next sections will couple hybrid moment closure with Yang-Mills field dynamics to give a hybrid formulation of Yang-Mills plasmas. Then, an application to ferromagnetic fluids with quadrupolar order will be presented.

\section{Coupling to the Yang-Mills field}
The preceding sections introduced the hybrid closure of the Yang-Mills Vlasov equation and investigated its geometric momentum map properties. This was done without considering the effects of the Yang-Mills fields. On the other hand, the YM Vlasov kinetic description first arose in the context of quark-gluon plasmas \cite{GiHoKu1983}, in which the field effects must be considered. This section shows how the hybrid closure combines with YM field dynamics. While this has mainly a pedagogical value, the implications of this construction in condensed matter systems are diverse. Indeed, the MHD limit of the whole Yang-Mills Vlasov system possesses the same Poisson bracket construction producing spin glass dynamics \cite{HoKu1988} and liquid crystal theories \cite{Ho2002}. Thus, motivated by this suggestive analogy, we shall proceed by formulating the hybrid theory of Yang-Mills plasmas and by applying its MHD limit to ferromagnetic hydrodynamics with quadrupolar order.

\rem{ %%%%%%%%%%%%%%%%%%%%%%%%%%
In condensed matter applications, one is not only interested in the order parameter micromotion. Another fundamental feature of complex fluids is the presence of defects \cite{KaEd}, such as line defects in the orientational order (\emph{disclinations}). Besides the well established topological approaches, one models disclinations by the use of an auxiliary variable known as \emph{disclination density}, which arises as the magnetic component of a YM field and obeys the same type of dynamics in either spin glasses or nematic liquid crystals. See \cite{Ho2002} for an explanatory example of how the disclination dynamics appears in liquid crystal dynamics. Accounting for the disclination density in the hybrid framework requires establishing the corresponding YM MHD formulation. In order to accomplish this task, we shall first construct the hybrid model for YM plasmas. Its MHD limit will be taken later upon using standard asymptotic methods.
} %%%%%%%%%%%%%%%%%%%%%%%%%%

\subsection{Hybrid description of Yang-Mills plasmas}
The YM Vlasov system is defined on the space
$\mathrm{Den}(T^*\Bbb{R}^3\times\mathfrak{o}^*)\times T^*\Omega^1(\Bbb{R}^3,\mathfrak{o})$,
where $\Omega^1(\Bbb{R}^3,\mathfrak{o})$ is the space of magnetic vector potentials. The Poisson bracket is given by the sum of a Vlasov term and (minus) a canonical one, involving the field variables $\bf (A,E)$:
\begin{equation}\label{PBminus}
\{F,K\}=\int \!f\left\{\frac{\delta F}{\delta f},\frac{\delta
K}{\delta f}\right\}_{1\!}\de^3 \mathbf{x}\,\de^3
\mathbf{p}\,\de\sigma-\left\{F,K\right\}{\bf (A,E)}
\end{equation}
where one defines \cite{GiHoKu1983}
\[
\left\{\frac{\delta F}{\delta f},\frac{\delta K}{\delta f}\right\}_1:=\left\{\frac{\delta F}{\delta f},\frac{\delta
K}{\delta f}\right\}
+
\left\langle\sigma,\left[\frac{\partial}{\partial\sigma}\frac{\delta
F}{\delta f},\frac{\partial}{\partial\sigma}\frac{\delta K}{\delta
f}\right]\right\rangle\,,
\]
and
\[
\{F,K\}{\bf (A,E)}=\int\!\left(\left\langle\frac{\delta F}{\delta {\bf A}},\frac{\delta
K}{\delta \mathbf{E}}\right\rangle-\left\langle\frac{\delta F}{\delta \mathbf{E}},\frac{\delta K}{\delta
\mathbf{A}}\right\rangle\right)\de^3 \mathbf{x}
\]
The minus sign in \eqref{PBminus} is consistent with the definition
of the electric part $\bf E$ of the field as (\emph{minus}) the momentum
variable conjugate to the vector potential $\bf A$ \cite{GiHoKu1983}. The YM
Vlasov Hamiltonian is given by \eqref{YMVHam}.

The new cold-plasma closure \eqref{Tenerife} generates a system on
$\mathfrak{X}^*(\Bbb{R}^3)\times \mathrm{Den}(\Bbb{R}^3\times\mathfrak{o}^*)\times T^*\Omega^1(\Bbb{R}^3,\mathfrak{o})$, where the space $\mathfrak{X}^*(\Bbb{R}^3)$ of one-form densities on physical space contains the fluid momentum variable.
The Hamiltonian \eqref{YMVHam} becomes
\begin{multline}\label{hybrid-ham1}
H=\frac12\int\!\rho\left|\frac{\mathbf{m}}{D}-\left\langle\sigma,\mathbf{A}\right\rangle\right|^2{\rm d}^3\mathbf{x}\,{\rm d}\sigma+\int \!D\,U(D)\,{\rm d}^3\mathbf{x}
-\int\! \rho \left\langle\sigma,A_0\right\rangle{\rm d}^3\mathbf{x}\,{\rm d}\sigma
\\
+\frac12\int\!\left(
\left|d^\mathbf{A} \mathbf{A}\right|^2+\left|\mathbf{E}\right|^2\right){\rm d}^3\mathbf{x}-\int\!\left\langle
\mathbf{E},d^\mathbf{A}A_0\right\rangle{\rm d}^3\mathbf{x}
\end{multline}
where the internal energy $D\,U(D)$ has been inserted again to account
for pressure effects. The new Poisson bracket in terms of the
variables $(D,\rho,\mathbf{m,A,E})$ reads
\begin{align}\nonumber
\{F,K\}=& \int \mathbf{m}\cdot
\left[\frac{\delta F}{\delta \mathbf{m}},\frac{\delta K}{\delta
\mathbf{m}}\right]_\mathfrak{\!X}{\rm d}^3\mathbf{x}+\int \rho\,\left\langle\sigma,\left[\frac{\partial}{\partial\sigma}\frac{\delta
F}{\delta \rho},\frac{\partial}{\partial\sigma}\frac{\delta K}{\delta \rho}\right]\right\rangle{\rm d}^3\mathbf{x}\,{\rm d}\sigma
-\{F,K\}(\mathbf{A,E})
\\
&- \int \rho\left(\pounds_{\delta F/\delta \mathbf{m}\,}\frac{\delta
K}{\delta \rho}-\pounds_{\delta K/\delta \mathbf{m}\,}\frac{\delta F}{\delta
\rho}\right){\rm d}^3\mathbf{x}\,{\rm d}\sigma
- \int D\left(\pounds_{\delta F/\delta \mathbf{m}\,}\frac{\delta
K}{\delta D}-\pounds_{\delta K/\delta \mathbf{m}\,}\frac{\delta F}{\delta
D}\right){\rm d}^3\mathbf{x}
\,,
\label{PB1}
\end{align}
where one recalls the constraint $D=\int\rho\,d\sigma$. This Poisson
bracket is evidently the sum of \eqref{PBwithD} and
$-\left\{F,G\right\}(\mathbf{A,E})$. The equations of motions follow naturally by inserting the Hamiltonian \eqref{hybrid-ham1} in the Poisson bracket \eqref{PB1}.

So far, the present construction yields the dynamics of the macroscopic fluid
momentum $\mathbf{m}(\mathbf{x},t)=\int\!\mathbf{p}\, f(\mathbf{x,p},\sigma,t)\,d^3\mathbf{p}\,d\sigma$, which is
not the variable commonly used in plasma physics. Indeed, a central
role is usually played by the equation of motion for the macroscopic
fluid velocity. At the particle level, the velocity
coordinate is given by the minimal coupling formula
$\boldsymbol{v}=\mathbf{p}-\left\langle\sigma,\mathbf{A}\right\rangle$, while we recall that in the
macroscopic fluid setting of CHD one writes
$\mathbf{u}=D^{-1\!}\left(\mathbf{m}-\left\langle G,\mathbf{A}\right\rangle\right)$, see equation \eqref{YM-velocity}. Although
the Hamiltonian \eqref{hybrid-ham1} suggests the velocity definition
$\mathbf{u}=D^{-1\!}\left(\mathbf{m}-\int\!\rho\left\langle\sigma,\mathbf{A}\right\rangle
d\sigma\right)$, a careful analysis requires to apply the
minimal coupling formula at the microscopic particle level, and to
define the averaged macroscopic fluid velocity only at a second
stage. Then,  the
hybrid equations can be written in terms of the macroscopic fluid
velocity, which is expressed by using moments of the Vlasov distribution
function.

In analogy with the preceding discussion, one starts with the
Poisson bracket \eqref{PBminus}
%\begin{equation}
%\{F,G\}=\int f\left\{\frac{\delta F}{\delta f},\frac{\delta
%G}{\delta f}\right\}_1-\{F,G\}(\mathbf{A,E})
%\end{equation}
and the YM Vlasov Hamiltonian \eqref{YMVHam}.
%\begin{equation}
%H=\frac12 \int
%f\left|\mathbf{p}-\left\langle\sigma,\mathbf{A}\right\rangle\right|^2-\int f
%\left\langle\sigma,A_0\right\rangle
%+\frac12\int\left( \left|d^\mathbf{A}
%\mathbf{A}\right|^2+\left|\mathbf{E}\right|^2\right)-\int\left\langle
%\mathbf{E},d^\mathbf{A}A_0\right\rangle
%\end{equation}
Then, in order to introduce minimal coupling, one utilizes
the well known relation
\begin{equation}
\boldsymbol{v}=\mathbf{p}-\langle\sigma,\mathbf{A}\rangle
\end{equation}
between momentum and velocity coordinates. Thus, one expresses the
system in terms of the distribution $\hat{f}(\mathbf{x},\boldsymbol{v},\sigma,t)$. The Poisson
bracket becomes
\begin{align}
\{F,K\}=& \int\hat{f} \left\{\frac{\delta F}{\delta \hat{f}},\frac{\delta
K}{\delta \hat{f}}\right\}_2- \{F,K\}(\mathbf{E,A}) + \int \hat{f}\left\langle\sigma,
\left(\frac{\partial}{\partial \boldsymbol{v}}\frac{\delta F}{\delta
\hat{f}}\cdot\frac{\delta K}{\delta \mathbf{E}}-\frac{\partial}{\partial
\boldsymbol{v}}\frac{\delta K}{\delta \hat{f}}\cdot\frac{\delta F}{\delta
\mathbf{E}}\right)\right\rangle
\label{YMVbracket}
\end{align}
where we define \cite{MoMaRa84}
\begin{multline*}
\left\{\frac{\delta F}{\delta \hat{f}},\,\frac{\delta K}{\delta
\hat{f}}\right\}_2:=\left\{\frac{\delta F}{\delta \hat{f}},\,\frac{\delta
K}{\delta \hat{f}}\right\}_1 +
\left\langle\sigma,\,\mathbf{B}\cdot\left(\frac{\partial}{\partial
\boldsymbol{v}}\frac{\delta F}{\delta \hat{f}}\times\frac{\partial}{\partial
\boldsymbol{v}}\frac{\delta K}{\delta \hat{f}}\right)\right\rangle
\\
+\left\langle\sigma,\left(\left[\mathbf{A}\cdot\frac{\partial}{\partial
\boldsymbol{v}}\frac{\delta F}{\delta
\hat{f}},\frac{\partial}{\partial\sigma}\frac{\delta K}{\delta
\hat{f}}\right]-\left[\mathbf{A}\cdot\frac{\partial}{\partial \boldsymbol{v}}\frac{\delta
K}{\delta \hat{f}},\frac{\partial}{\partial\sigma}\frac{\delta F}{\delta
\hat{f}}\right]\right)\right\rangle
\end{multline*}
Once minimal coupling has been introduced, we are ready to derive the hybrid model for Yang-Mills plasmas by adopting the hybrid Vlasov closure
\begin{equation}
\hat{f}(\mathbf{x},\boldsymbol{v},\sigma,t)=\rho(\mathbf{x},\sigma,t)\
\delta\!\left(\boldsymbol{v}-\frac{\mathbf{M}(\mathbf{x},t)}{D(\mathbf{x},t)}\right) \,.
\end{equation}
Upon inserting the chain rule relation
\begin{equation}
\frac{\delta F}{\delta \hat{f}}=\frac{\delta F}{\delta D}+\frac{\delta
F}{\delta \rho}+\boldsymbol{v}\cdot\frac{\delta F}{\delta \mathbf{M}}
\,,
\end{equation}
in \eqref{YMVbracket}, we obtain  the hybrid Poisson
bracket
\begin{align}\nonumber
\{F,K\}=& \int \mathbf{M}\cdot \left[\frac{\delta F}{\delta \mathbf{M}},\frac{\delta
K}{\delta \mathbf{M}}\right]_\mathfrak{\!X}{\rm d}^3\mathbf{x}- \int
D\left(\pounds_{\delta F/\delta \mathbf{M}\,}\frac{\delta K}{\delta
D}-\pounds_{\delta K/\delta \mathbf{M}\,}\frac{\delta F}{\delta D}\right){\rm
d}q
\\ \nonumber
&- \int\!\rho\left(\pounds_{\delta F/\delta \mathbf{M}\,}\frac{\delta
K}{\delta \rho}-\pounds_{\delta K/\delta \mathbf{M}\,}\frac{\delta F}{\delta
\rho}\right){\rm d}^3\mathbf{x}\,{\rm d}\sigma-\{F,K\}(\mathbf{E,A})
\\ \nonumber
&+\int\!
\rho\left\langle\sigma,\left[\frac{\partial}{\partial\sigma}\frac{\delta
F}{\delta \rho},\frac{\partial}{\partial\sigma}\frac{\delta
K}{\delta \rho}\right]\right\rangle{\rm d}^3\mathbf{x}\,{\rm d}\sigma +
\int\!\rho\left\langle\sigma,\,\mathbf{B}\cdot\frac{\delta F}{\delta
\mathbf{M}}\times\frac{\delta K}{\delta \mathbf{M}}\right\rangle \de^3\mathbf{x}
\,\de\sigma
\\ \nonumber
&+\int\!\rho\left\langle\sigma,\left[\mathbf{A}\cdot\frac{\delta
F}{\delta \mathbf{M}},\frac{\partial}{\partial\sigma}\frac{\delta K}{\delta
\rho}\right]-\left[\mathbf{A}\cdot\frac{\delta K}{\delta
\mathbf{M}},\frac{\partial}{\partial\sigma}\frac{\delta F}{\delta
\rho}\right]\right\rangle\de\sigma\,\de \mathbf{x}
\\
&+ \int\!\rho\left\langle\sigma,\frac{\delta F}{\delta
\mathbf{M}}\cdot\frac{\delta K}{\delta \mathbf{E}}-\frac{\delta K}{\delta
\mathbf{M}}\cdot\frac{\delta F}{\delta \mathbf{E}}\right\rangle\de\sigma\,\de \mathbf{x}
\label{PB2}
\end{align}
In the multi-fluid case, the above bracket represents the hybrid
Yang-Mills version of the Spencer-Kaufman bracket for
multicomponent magnetized plasmas \cite{Spencer,SpKa}.

The  Hamiltonian \eqref{YMVHam} now takes  the hybrid form
\begin{equation}
H=\frac12 \int \frac{\left|\mathbf{M}\right|^2}{D}-\int \rho
\left\langle\sigma,A_0\right\rangle+\int D\,U(D)
\\
+\frac12\int\left( \left|d^\mathbf{A}
\mathbf{A}\right|^2+\left|\mathbf{E}\right|^2\right){+}\int\left\langle
\mathbf{E},d^\mathbf{A}A_0\right\rangle
\end{equation}
so that Hamilton's equations
\rem{ %%%%%%%%%%%%%%%%%%%%%%%%%%%%%%%%
\begin{align*}\nonumber
\frac{\partial D}{\partial t}+&{\rm div\!}\left(D\,\frac{\delta
H}{\delta \mathbf{M}}\right)=0
\\
\frac{\partial \mathbf{M}}{\partial t}+&\pounds_{\delta H/\delta
\mathbf{M}\,}\mathbf{M}=-\left\langle\int\rho\,\sigma\,\de\sigma,\left(\frac{\delta
H}{\delta \mathbf{E}}+\frac{\delta H}{\delta \mathbf{M}}\times
\mathbf{B}\right)\right\rangle-\int\rho\left\langle\sigma,\left[
\frac{\partial}{\partial\sigma}\frac{\delta H}{\delta
\rho},\mathbf{A}\right]\right\rangle
\\
&\hspace{7.5cm}-\int\!\rho\,\nabla \frac{\delta H}{\delta
\rho}\,\de\sigma-D\nabla\frac{\delta H}{\delta D}
\\
\nonumber \frac{\partial\rho}{\partial t}+&{\rm
div\!}\left(\rho\,\frac{\delta H}{\delta
\mathbf{M}}\right)=-\left\langle\sigma,\left[\frac{\partial \rho}{\partial
\sigma},\left(\frac{\partial}{\partial \sigma}\frac{\delta H}{\delta
\rho}-\mathbf{A}\cdot\frac{\delta H}{\delta \mathbf{M}}\right)\right]\right\rangle
\end{align*}
} %%%%%%%%%%%%%%%%%%%%%%%%%%%%%%%%
become simply
\begin{align*}\nonumber
&\frac{\partial D}{\partial t}+{\rm div\!}\left(D\mathbf{u}\right)=0
\\
&D\!\left(\frac{\partial \mathbf{u}}{\partial t}+\mathbf{u}\cdot\nabla \mathbf{u}\right)=-\,\nabla{\sf
p}-\left\langle\int\rho\,\sigma\,\de\sigma,\,\mathbf{E}+\mathbf{u}\times
\mathbf{B}\right\rangle
\\
\nonumber &\frac{\partial\rho}{\partial t}+{\rm
div\!}\left(\rho\,\mathbf{u}\right)=\left\langle\sigma,\left[\frac{\partial
\rho}{\partial \sigma},\,A_0+\mathbf{A\cdot u}\right]\right\rangle
,
\end{align*}
which are to be accompanied by Gau{\ss} Law
\[
\frac{\delta H}{\delta A_0} =0=\operatorname{div}^\mathbf{A}
\mathbf{E}{+}\int\rho\,\sigma \,\de\sigma
\]
and the field equations
\begin{align*}
\frac{\partial \mathbf{E}}{\partial t}&=\frac{\delta H}{\delta
\mathbf{A}}-\frac{\delta H}{\delta
\mathbf{M}}\int\!\rho\,\sigma\,\de\sigma=\operatorname{curl}^\mathbf{A}(d^\mathbf{A}
\mathbf{A}){+}\operatorname{ad}^*_{A_0}\mathbf{E}-\mathbf{u}\!\int\!\rho\,\sigma\,\de\sigma
\\
\frac{\partial \mathbf{A}}{\partial t}&=-\frac{\delta H}{\delta \mathbf{E}}=-\mathbf{E}-d^\mathbf{A} A_0
\end{align*}
The first moment of the $\rho-$equation yields the continuity equation in the covariant form
\[
\left(\frac{\partial}{\partial t}+\operatorname{ad}^*_{A_0}\right)\int\rho\,\sigma
\,\de\sigma+\operatorname{div}^\mathbf{A}\!\left(\mathbf{u}\!\int\!\rho\,\sigma
\,\de\sigma\right)=0
\,.
\]
Notice that these equations contain precisely the same amount of physical information as the chromohydrodynamics equations \eqref{YM-velocity}, \eqref{YM-fields} and \eqref{YM-transport} (see also \cite{GiHoKu1983}) because the dynamics of the Yang-Mills charge density $G=\int\!\rho\,\sigma\,\de\sigma$ does not involve any higher-order moment of $\rho$. 

%%%%%%%%%%%%%%%%%%%%%%%%%%%%%%%%%%%%%%%%%

{
\begin{remark}[Hybrid Yang-Mills MHD]\label{MHDsection}
The MHD limit of chromohydrodynamics (Yang-Mills MHD, or YM-MHD) is known to possess many condensed matter applications. For example, the equations of spin glasses were shown \cite{HoKu1988} to possess the same Hamiltonian structure as Yang-Mills MHD. Thus, one is interested in finding the hybrid analogue of YM-MHD. This task can be easily accomplished by following precisely the same steps as in \cite{HoKu1988}, upon recalling that the total YM charge is now written as $G(\rho)= \int\!\rho\,\sigma\de\sigma$. 
%%%%%%%%%%%%%%%%%%%%%%%%%%%%%%%%%%%%%%%%%%%%%%%%%%%%%%%%%%%%%%%%%%%%%%%
YM-MHD shares many analogies with ferromagnetic hydrodynamics \cite{GiHoKu1983}, which is generalized in the next section to systems with quadrupolar order.
\end{remark}
}

\subsection{Ferromagnetic fluids with quadrupolar order}
The equations of ferromagnetic hydrodynamics (FHD) possess a well known Hamiltonian structure, which was discovered and investigated in \cite{HoKu1998bis}. More recently, a Vlasov equation of the same type as in \cite{GiHoKu1983} has been presented in relation to magnetized plasmas \cite{BrMaZaErMa2008,MaMo2011}. In this section, we present an intermediate hybrid formulation of ferromagnetic continua, which illustrates a direct application of the hybrid approach to the case of quadrupolar interactions. {\color{black}For the purpose of simplicity, we shall consider incompressible fluid flows and the fluid momentum $\mathbf{m}$ will be denoted by $\mathbf{u}$.}

Based on the work of Holm and Kupershmidt \cite{HoKu1998bis}, it is easy to write the hybrid total energy for ferromagnetic (incompressible) hydrodynamics with quadrupolar order (cf. \cite{AnGr}):
\begin{equation}\label{FHD-Ham}
H=\frac12 \int \left|\mathbf{u}\right|^2\,\de^3\mathbf{x}
+
\frac1{8\pi}\int\! \left|\mathbf{B}-4\pi \boldsymbol{\mathcal{M}}(\rho)\right|^2\de\sigma\,\de^3\mathbf{x}+\frac\mu2\!\int\!\operatorname{Tr}\!\left(Q^2(\rho)\right)\de^3\mathbf{x}
%+\int\! D\,U(D)\,\de^3\mathbf{x}
\,,
\end{equation}
where the tensor $Q$ is defined in \eqref{Q-spintensor} and  $\boldsymbol\sigma\in\mathfrak{so}(3)\simeq\Bbb{R}^3$ is the particle magnetic spin interacting with the magnetic induction field $\mathbf{B}$.
Evidently, when $\mu=0$, the above expression recovers the incompressible version of the fluid Hamiltonian in \cite{HoKu1998bis}. 
{
Upon restricting to incompressible flows, the hybrid equations of motion arise from the Poisson bracket 
\begin{align}\nonumber
\{F,K\}(\mathbf{u},\rho,B)=& \int \mathbf{u}\cdot\operatorname{curl}\!
\left(\frac{\delta F}{\delta \mathbf{u}}\times\frac{\delta K}{\delta
\mathbf{u}}\right){\rm d}^3\mathbf{x}+\int \! \rho\left\langle\boldsymbol\sigma,\left[\frac{\partial}{\partial\boldsymbol\sigma}\frac{\delta
F}{\delta \rho},\frac{\partial}{\partial\boldsymbol\sigma}\frac{\delta K}{\delta \rho}\right]\right\rangle{\rm d}^3\mathbf{x}\,{\rm d}\boldsymbol\sigma
\\
\nonumber
&- \int \rho\left(\pounds_{\delta F/\delta \mathbf{u}\,}\frac{\delta
K}{\delta \rho}-\pounds_{\delta K/\delta \mathbf{u}\,}\frac{\delta F}{\delta
\rho}\right){\rm d}^3\mathbf{x}\,{\rm d}\sigma
\\
&- \int \left\langle\mathbf{B},\pounds_{\delta F/\delta \mathbf{u}\,}\frac{\delta
G}{\delta \mathbf{B}}-\pounds_{\delta K/\delta \mathbf{u}\,}\frac{\delta F}{\delta
\mathbf{B}}\right\rangle{\rm d}^3\mathbf{x}
\,.
\label{PB-FHD}
\end{align}
which is derived from the Hamiltonian structure in \cite{GiHoKu1983} upon replacing $\boldsymbol{\mathcal{M}}=\int{\boldsymbol\sigma}\,\rho\,\de^3{\boldsymbol\sigma}$. For the Lie algebra $\mathfrak{so}(3)\simeq\Bbb{R}^3$, the Lie bracket $[\cdot,\cdot]$ above is given by the cross product, i.e. $[\boldsymbol\xi,\boldsymbol\nu]=\boldsymbol\xi\times\boldsymbol\nu$. Also, all the angle brackets $\langle\cdot,\cdot\rangle$ above are simply dot products. 
\begin{remark}[Ferromagnetic hydrodynamics v.s. Yang-Mills MHD]
Notice that the pairings $\langle\cdot,\cdot\rangle$ appearing in \eqref{PB-FHD} allow for the more general case when the magnetic induction $\mathbf{B}$ is replaced by a Yang-Mills field taking values in the Lia algebra $\mathfrak{o}$, whose dual $\mathfrak{o}^*$ contains the Yang-Mills charge $\sigma$. This produces precisely the Poisson bracket of hybrid YM-MHD. However, for ferromagnetic hydrodynamics, one has $\boldsymbol\sigma\in\mathfrak{so}(3)^*\simeq\Bbb{R}^3$ while the magnetic field is simply a two-form $\mathbf{B}\cdot\de\mathbf{S}\in\Omega^2(\Bbb{R}^3)$ (here, $\de\mathbf{S}$ denotes an oriented surface element in $\Bbb{R}^3$), so that the angle brackets $\langle\cdot,\cdot\rangle$ in \eqref{PB-FHD} reduce to dot product.
\end{remark}
The equations of motion} read
\begin{align*}
&\frac{\partial\mathbf{u}}{\partial t}+(\mathbf{u}\cdot\nabla)\mathbf{u}+\nabla p=-\int\!\rho\,\nabla
\frac{\delta H}{\delta \rho}\,\de\boldsymbol\sigma-\mathbf{B}\times\operatorname{curl}\frac{\delta H}{\delta
\mathbf{B}}
\\
\nonumber &\frac{\partial\rho}{\partial t}+\mathbf{u}\cdot\nabla\rho=-g\,\boldsymbol\sigma\cdot\frac{\partial \rho}{\partial
\boldsymbol\sigma}\times\frac{\partial}{\partial \boldsymbol\sigma}\frac{\delta H}{\delta
\rho}
\\
&\frac{\partial\mathbf{B}}{\partial
t}+\nabla\times(\mathbf{B\times u)}=0
\end{align*}
where the constant gyromagnetic ratio $g$ has been inserted in the $\rho$-equation for consistency.
Upon using the hybrid Hamiltonian \eqref{FHD-Ham}, 
\rem{ %%%%%%%%%%%%%%%%%%%%%%%%%%%%%%%
we compute
\begin{align*}
\frac{\delta H}{\delta
\mathbf{m}}=&\frac{\mathbf{m}}D=:\mathbf{u}\\
\frac{\delta H}{\delta
D}=&-\frac{|\mathbf{m}|^2}{D^2}-\frac1{8\pi}\left(\frac{|\mathbf{B}|^2}{D}-16\pi^2\!\int\!\rho\,|\boldsymbol\sigma|^2\,\de\boldsymbol\sigma\right)
+U(D)+D\,U'(D)
\\
\frac{\delta H}{\delta \rho}=&\frac1{8\pi D}\left|\mathbf{B}-4\pi D
\boldsymbol\sigma\right|^2
\\
\frac{\delta H}{\delta \mathbf{B}}=&\frac1{4\pi}\left(\mathbf{B}-4\pi\!\int\!\rho\,\boldsymbol\sigma\,\de\boldsymbol\sigma\right)=:\frac1{4\pi}\boldsymbol{\mathcal{H}} \,,
\end{align*}
where $\mathbf{u}$ is the fluid velocity. Thus, upon 
}     %%%%%%%%%%%%%%%%%%%%%%%%%%%%%%%
redefining the pressure appropriately yields
\rem{ %%%%%%%%%%%%%%%%%%%%%%%%%%%%%%%
\begin{align*}
-D\nabla\frac{\delta H}{\delta
D}-\int\rho\nabla \frac{\delta H}{\delta
\rho}&=D\nabla\left(\frac{M^2}{D^2}\right)-\nabla{\sf p}-2\pi D\,\nabla\!\int\!\rho\,\sigma^2\,\de\sigma
-\int\!\rho\,\nabla\!\left(2\pi D\sigma^2- B\cdot\sigma\right)\de\sigma
\\
&=
D\,\nabla\!\left(\frac{M^2}{D^2}\right)-\nabla{\sf p}-2\pi \nabla\!\left(D\!\int\!\rho\,\sigma^2\,\de\sigma\right)
+
\nabla B\cdot\int\rho\,\sigma\, \de\sigma
 \,,
\end{align*}
so that 
}  %%%%%%%%%%%%%%%%%%%%%%%%%%%%%%%
the hybrid equations
\begin{align*}
&\frac{\partial\mathbf{u}}{\partial t}+(\mathbf{u}\cdot\nabla)\mathbf{u}+\nabla{p}=
\nabla \mathbf{B}\cdot\int\rho\,\boldsymbol\sigma\,\de\boldsymbol\sigma-\frac1{4\pi}\mathbf{B}\times\nabla\times\!\left(\mathbf{B}-4\pi\!\int\rho\,\boldsymbol\sigma\,\de\boldsymbol\sigma\right)
\\
\nonumber 
& \frac{\partial\rho}{\partial t}+{\rm
div\!}\left(\rho\,\mathbf{u}\right)=g \left(\mathbf{B}-4\pi\boldsymbol{\mathcal{M}}-2\mu Q\boldsymbol\sigma\right)\cdot\boldsymbol\sigma \times\frac{\partial \rho}{\partial
\boldsymbol\sigma}
\\
&\frac{\partial\mathbf{B}}{\partial
t}+\nabla\times(\mathbf{B\times u)}=0
\end{align*}
which evidently reduce to the incompressible form of the fluid equations in \cite{HoKu1998bis} by simply setting $\mu=0$. 
\rem{ %%%%%%%%%%%%%%%%%%%%%%%%%%%%
The velocity equation reads as
\[
D\!\left(\frac{\partial V}{\partial t}+\left(V\cdot\nabla\right)V\right)+\nabla{\sf p}=\nabla B\cdot
\int\!\rho\,\sigma\de\sigma-2\pi \nabla\!\left(D\,\!\int\!\rho\,\sigma^2\,\de\sigma\right)-\frac1{4\pi}B\times\operatorname{curl}\mathcal{H}\,.
\]

It may be advantageous to express the hybrid equations for FHD in terms of the quantity $\varphi=\rho/D$. These are readily written as
\begin{align*}
&D\!\left(\frac{\partial \mathbf{u}}{\partial t}+\left(\mathbf{u}\cdot\nabla\right)\mathbf{u}\right)+\nabla{\sf p}=D\nabla \mathbf{B}\cdot
\int\!\varphi\,\boldsymbol\sigma\de\boldsymbol\sigma-2\pi \nabla\!\left(D^2\!\int\!\varphi\,|\boldsymbol\sigma|^2\,\de\boldsymbol\sigma\right)-\frac1{4\pi}\mathbf{B}\times\operatorname{curl}\boldsymbol{\mathcal{H}}
\\
\nonumber 
&\frac{\partial\varphi}{\partial t}+\mathbf{u}\cdot\nabla\varphi=-g\,\mathbf{B}\cdot\boldsymbol\sigma \times\frac{\partial \varphi}{\partial
\boldsymbol\sigma}
\\
&\frac{\partial D}{\partial t}+{\rm div\!}\left(D\,\mathbf{u}\right)=0 \,,\qquad\ \frac{\partial \mathbf{B}}{\partial
t}-\operatorname{curl}\left(\mathbf{u}\times \mathbf{B}\right)=0\,,
\end{align*}
}   %%%%%%%%%%%%%%%%%%%%%%%%%%%%
We recall that the quantity $C(\rho)=\int \!\Lambda(\rho)\,\de\sigma\,\de \mathbf{x}$ determines a whole family of Casimir invariants for the corresponding Poisson  bracket \eqref{PB-FHD}. Again, these can be used to apply the energy-Casimir method by following \cite{HoMaRaWe1985}.
Also, the magnetic helicity $\int\!\mathbf{A}\cdot \mathbf{B}\,\de^3 \mathbf{x}$ (where $\mathbf{B}=\operatorname{curl}\mathbf{A}$) is another Casimir for the Poisson structure of incompressible FHD, when the latter is expressed in terms of the vector potential $\mathbf{A}\cdot\de\mathbf{x}\in\Omega^1(\Bbb{R}^3)$. While stability is left for future work, the next section extends the previous models to account for systems with symmetry breaking. More particularly, the case of liquid crystal flows will be considered.

\section{Hybrid models for systems with symmetry breaking}

In soft matter systems one often deals with broken symmetries. As
briefly mentioned in the Introduction, this happens when the single
particle dynamics is not entirely symmetric under its underlying
configuration Lie group $\mathcal{O}$. Usually the particle dynamics
allows for a residual symmetry $\mathcal{P}$, which is normally used
to describe the particle micromotion on the symmetry-reduced phase
space $T^*\mathcal{O/P}\simeq\mathfrak{o}^*\times(\mathcal{O/P})$. See e.g. \cite{GayBalmazTronci, Ho2002} for how the coset space emerges from the reduction of systems with symmetry breaking.
Here, the coset space $M=\mathcal{O/P}$ is the
\emph{order parameter space}, while $\mathcal{O}$ is sometimes
called the \emph{broken symmetry group}. The reduction theory for
systems with broken symmetry has been developed in \cite{GayBRatiu,GayBalmazTronci, Ho2002}, to which
we address for further details. In the remaining sections, the previous results on hybrid models will be extended to systems with broken symmetry. A hybrid formulation of liquid crystal dynamics will be presented and compared to Eringen's micropolar theory. Applications to cubatic liquid crystals with multipolar order will be proposed.

\subsection{Yang-Mills-Vlasov equation with symmetry breaking}

The YM Vlasov equation for systems with broken symmetry is written
immediately by simply replacing the Lie-Poisson bracket on
$\mathfrak{o}^*$ in \eqref{YMV} by the natural generalized
Lie-Poisson bracket on $\mathfrak{o}^*\times M$ \cite{GayBalmazTronci}. This leads
to the following form of the YM Vlasov equation
\begin{equation}\label{YMVwbs}
\frac{\partial f}{\partial t}+\left\{f,\frac{\delta H}{\delta
f}\right\}+\left\langle\sigma,\left[\frac{\partial f}{\partial
\sigma},\frac{\partial}{\partial \sigma}\frac{\delta H}{\delta
f}\right]\right\rangle - \left\langle \frac{\partial f}{\partial
n},\left(\frac{\partial}{\partial \sigma}\frac{\delta H}{\delta
f}\right)_{\!\!M} n\right\rangle+\left\langle
\frac{\partial}{\partial n}\frac{\delta H}{\delta
f},\left(\frac{\partial f}{\partial \sigma}\right)_{\!\!M}
n\right\rangle=0
\end{equation}
where $n\in M$ is the order parameter coordinate and the notation
$\xi_M\, n$ stands for the infinitesimal action of $\mathcal{O}$ on
$M$. Here the angle bracket $\langle\cdot,\cdot\rangle$ denotes
either the pairing between the Lie algebra and its dual and the
pairing between vectors and covectors on $M$. 

In order to obtain the fluid description of complex fluids with
broken symmetry, one considers the following Vlasov distribution
function $f(\mathbf{x,p},n,\sigma,t)$:
\[
f(\mathbf{x,p},\sigma,n,t)=D(\mathbf{x},t)\ \delta\!\left(n-{\sf
n}(\mathbf{x},t)\right)\,\delta\!\left(\mathbf{p}-\frac{\mathbf{m}(\mathbf{x},t)}{D(\mathbf{x},t)}\right)\delta\!\left(\sigma-\frac{G(\mathbf{x},t)}{D(\mathbf{x},t)}\right)
\]
In a typical situation one has $M=S^2\subset\Bbb{R}^3$, that is the
order parameter space $M$ is the unit sphere $S^2$ and points $\boldsymbol{n}\in
S^2$ on the sphere are identified with unit vectors in $\Bbb{R}^3$,
so that $\xi_M\,\boldsymbol{n}=\boldsymbol\xi\times \boldsymbol{n}$. 
%This is the situation, for example, in spin glass dynamics and other ferromagnetic media. 
The fluid
equations {\color{black}for an arbitrary Hamiltonian $H(\mathbf{m},G,\mathsf{n},D)$} read as \cite{Ho2002,GayBRatiu,IsKoPe}
\begin{align}
\nonumber &\left(\frac{\partial }{\partial
t}+\pounds_{\textstyle\frac{\delta H}{\delta \mathbf{m}}}\right)D =0
\\
\nonumber &\left(\frac{\partial }{\partial
t}+\pounds_{\textstyle\frac{\delta H}{\delta \mathbf{m}}}\right)G ={-}
\operatorname{ad}^*_{\frac{\delta H}{\delta
G}}G+\mathsf{n}\diamond\frac{\delta H}{\delta \mathsf{n}}
\\
\nonumber &\left(\frac{\partial }{\partial
t}+\pounds_{\textstyle\frac{\delta H}{\delta \mathbf{m}}}- \left(\frac{\delta
H}{\delta G}\right)_{\!\!M}\right){\sf n} =0
\\
&\left(\frac{\partial}{\partial t}+\pounds_{\textstyle\frac{\delta
H}{\delta \mathbf{m}}}\right)\mathbf{m} = -D\,\nabla\frac{\delta H}{\delta D} -
\left\langle G,\nabla\frac{\delta H}{\delta
G}\right\rangle+\left\langle\frac{\delta H}{\delta
\mathsf{n}},\nabla \mathsf{n}\right\rangle
\label{CHDwsb}
\end{align}
where one defines the diamond operator $\diamond$ as $\left\langle
n\diamond \kappa,\xi\right\rangle:=\left\langle
\kappa,\xi_M\,n\right\rangle$, for any covector $\kappa\in T^*M$. Evidently, the quantity $n\diamond \kappa$ coincides with the momentum map for cotangent-lifted actions \cite{MaRa,HoScSt}.
The Poisson bracket {\color{black}corresponding to \eqref{CHDwsb}} is
\begin{align}\nonumber
\{F,K\}(\mathbf{m},D,G,\mathsf{n})=&  \int \mathbf{m}\cdot
\left[\frac{\delta F}{\delta \mathbf{m}},\frac{\delta K}{\delta
\mathbf{m}}\right]_\mathfrak{\!X}{\rm d}^3\mathbf{x}+\int \left\langle
G,\left[\frac{\delta F}{\delta G},\frac{\delta K}{\delta
G}\right]\right\rangle{\rm d}^3\mathbf{x}
\\ \nonumber
&- \int D\left(\pounds_{\delta F/\delta \mathbf{m}\,}\frac{\delta K}{\delta
D}-\pounds_{\delta K/\delta \mathbf{m}\,}\frac{\delta F}{\delta D}\right){\rm
d}^3\mathbf{x}
\\
\nonumber &- \int \left\langle G,\left.\pounds_{\delta F/\delta
\mathbf{m}\,}\frac{\delta K}{\delta G}-\pounds_{\delta K/\delta
\mathbf{m}\,}\frac{\delta F}{\delta G}\right.\right\rangle{\rm d}^3\mathbf{x}
\\
&+ \int \left(\left\langle\frac{\delta F}{\delta
\mathsf{n}},\left(\left(\frac{\delta K}{\delta G}\right)_{\!\!M}
-\pounds_{\textstyle\frac{\delta K}{\delta
\mathbf{m}}}\right)\mathsf{n}\right\rangle-\left\langle \frac{\delta
K}{\delta \mathsf{n}},\left(\left(\frac{\delta F}{\delta
G}\right)_{\!\!M}-\pounds_{\!\textstyle\frac{\delta F}{\delta
\mathbf{m}}}\right) \mathsf{n}\right\rangle\right){\rm d}^3\mathbf{x} \,,
\end{align}
This framework is particularly suitable for liquid crystals (the director { field} notation is $\mathsf{n}=\mathbf{n(x,}t)$), which will be the guiding example in this last section. Convenient Poisson bracket approaches to several types of liquid crystal dynamics were developed in \cite{Lubensky2003,DzVo,VoKa,IsKoPe}. For example, the dynamics of uniaxial nematic liquid crystals is governed by the equations \eqref{CHDwsb} and the Hamiltonian
\begin{equation}\label{LiqXalHamiltonian}
H=\frac12\int \frac{\left|\mathbf{m}\right|^2}{D}\,\de^3 \mathbf{x}+\frac1{2J}\int\left|\mathbf{G}\right|^2\,\de^3\mathbf{x}+\int\! D\,F(D^{-1},{\bf n},\nabla{\bf n})\,\de^3\mathbf{x}\,,
\end{equation}
where
\begin{equation}\label{FrankEnergy}
D\,F(D^{-1},{\bf n},\nabla{\bf n})=K_1(\operatorname{div}\mathbf{n})^2+(\mathbf{n}\cdot\nabla\times\mathbf{n})\left(K_2(\mathbf{n}\cdot\nabla\times\mathbf{n})+K_2'\right)+K_3|\mathbf{n}\times\nabla\times\mathbf{n}|^2
\end{equation}
is the Frank energy \cite{Ch1992,deGePr1993}. 

The presence {\color{black}in \eqref{FrankEnergy}} of gradients of the order parameter field reflects the elastic properties of the liquid crystal material. In more generality, the emergence of these gradients is responsible for the failure of kinetic mean field approaches to complex fluid dynamics. Indeed, while mean field approaches are well known to apply to equilibrium situations, their employment in dynamical problems present several difficulties. An interesting attempt to overcome these difficulties is given by the following mesoscopic closure developed in \cite{BlEhMu91,BlEhMu92}:
\[
f(\mathbf{x,p},\sigma,n,t)=\rho(\mathbf{x},n,t)\ \delta\!\left(\mathbf{p}-\frac{\mathbf{M}(\mathbf{x},n,t)}{\rho(\mathbf{x},n,t)}\right)\delta\!\left(\sigma-\frac{\Xi(\mathbf{x},n,t)}{\rho(\mathbf{x},n,t)}\right),
\]
where all hydrodynamical variables are defined on the extended configuration space $\Bbb{R}^3\times M$. Thus, the microscopic information on the order parameter space $M$ is retained by all dynamical variables. However, the resulting equations present several difficulties due to the high dimensionality of the space $\Bbb{R}^3\times M$, which makes the dynamics difficult to solve even by using advanced numerical methods.

The next sections show how these difficulties can be overcome by hybrid models, in which the macroscopic order parameter field is replaced by a position-dependent statistical distribution on the order parameter space $M$.

\subsection{Hybrid closure for systems with broken symmetry}

Recent computer simulations of cubatic liquid crystal phases \cite{DuDeMaWi,BaStTo} and other complex fluids with multipolar order raise the necessity of accounting for the effects of correlations higher than second order. Previous sections showed how, for magnetic media, these correlation effects are easily taken into account by the hybrid approach. The following sections address the problem of extending this approach to systems with broken symmetry, with special emphasis on liquid crystal dynamics.

A hybrid closure of the YM Vlasov equation for $f(\mathbf{x,p},\sigma,n)$ can be easily obtained upon treating both coordinates $(\sigma,n)$ at the same level, thereby obtaining the closure
\[
f(\mathbf{x,p},\sigma,n,t)=\rho(\mathbf{x},\sigma,n,t)\,\delta\!\left(\mathbf{p}-\frac{\mathbf{m}(\mathbf{x},t)}{D(\mathbf{x},t)}\right)\,,
\]
with $D(\mathbf{x})=\int\!\rho(\mathbf{x},\sigma,n)\,\de\sigma\de n$. The above closure applies to nematic systems in which different molecules at the same point possess different orientation and angular momentum. This is a rather general situation. 
\rem{ %%%%%%%%%%%%%%%%%%%%%%%%%%%%%%%%%%
{\color{black}An equivalent} description is obtained upon introducing the variable $\varphi=\rho/D$, with $\int\!\varphi\,\de\sigma\,\de n=1$. 
Then, the total energy \eqref{LiqXalHamiltonian} is then rewritten as
\begin{equation}
H=\frac12\int \frac{\left|M\right|^2}{D}\,\de \mathbf{x}+\frac1{2J}\int \!D^2\,\varphi\left|\sigma\right|^2\,\de \mathbf{x}+\int\! D\,F(D^{-1},\varphi,\nabla\varphi)\,\de \mathbf{x}\,,
\end{equation}
where the Frank energy is now written as
\begin{multline}
F=K_1\left|\int\! \boldsymbol{n}\cdot\nabla\varphi\,\de\sigma\,\de \boldsymbol{n}\right|^{2}+K_2\left|\int\!\varphi(\mathbf{x},\sigma',\boldsymbol{n}')\ \boldsymbol{n}'\cdot \boldsymbol{n}\times\nabla\varphi(\mathbf{x},\sigma,\boldsymbol{n})\ \de\sigma'\,\de \boldsymbol{n}' \,\de\sigma\,\de \boldsymbol{n}\right|^{2}
\\
+K_3\left|\int\!\varphi(\mathbf{x},\sigma',\boldsymbol{n}')\ \boldsymbol{n}'\times \boldsymbol{n}\times\nabla\varphi(\mathbf{x},\sigma,\boldsymbol{n})\ \de\sigma'\,\de \boldsymbol{n}' \,\de\sigma\,\de \boldsymbol{n}\right|^{2}\,.
\end{multline}
Then, the usual complex fluid equations can be recovered by the singular expression
\[
\varphi(\mathbf{x},\sigma,{n},t)=\delta\!\left(\sigma-\frac{G(\mathbf{x},t)}{D(\mathbf{x},t)}\right)\,\delta({n}-{\sf n}(\mathbf{x},t))\,.
\]
} %%%%%%%%%%%%%%%%%%%%%%%%%%%%%%%%%%
However, the position-dependent distribution $\rho(\mathbf{x},\sigma,n,t)$ still depends on a high number of variables and thus it is not easy to handle. Thus, although the above hybrid closure may apply to nematic liquid crystal dynamics, it is more convenient to make simplifying assumptions.

In the search for a hybrid description of complex fluids with
symmetry breaking, one can seek an equation for a probability
distribution on the space $\Bbb{R}^3\times M$. This is precisely the same approach that is followed, for example, in ordinary Smoluchowsky theories, where such a distribution obeys a diffusion-type kinetic equation in the overdamped limit {\cite{DoEd1988}}. {\color{black}On the other hand, in this paper we consider systems with non-negligible inertial effects, so that the following results differ from the predictions of the Smoluchowsky approach.} In the present setting, we shall consider the cold plasma closure
\[
f(\mathbf{x,p},\sigma,n,t)=D(\mathbf{x},t)\,\varphi(\mathbf{x},n,t)\
\delta\!\left(\mathbf{p}-\frac{\mathbf{m}(\mathbf{x},t)}{D(\mathbf{x},t)}\right)
\delta\!\left(\sigma-\frac{G(\mathbf{x},t)}{D(\mathbf{x},t)}\right)
\,,
\]
where  $\int \varphi(\mathbf{x},n,t)\,\de n=1$. This closure applies to situations in which different molecules at the same point possess different orientations and the same angular momentum. Thus, at each point in space one can have several rod-like molecules rotating in a moving plane with different orientations at all times. The resulting hybrid model possesses a Hamiltonian structure, which can be obtained by the same methods that were presented in the first half of this paper. One obtains the following Poisson bracket 
\begin{align}\nonumber
\{F,K\}(\mathbf{m},D,G,\varphi)=& \int \mathbf{m}\cdot \left[\frac{\delta F}{\delta
\mathbf{m}},\frac{\delta K}{\delta \mathbf{m}}\right]_\mathfrak{\!X}{\rm d}^3\mathbf{x}+\int
\left\langle G,\left[\frac{\delta F}{\delta G},\frac{\delta
K}{\delta G}\right]\right\rangle{\rm d}^3\mathbf{x}
\\ \nonumber
&- \int D\left(\pounds_{\delta F/\delta \mathbf{m}\,}\frac{\delta K}{\delta
D}-\pounds_{\delta H/\delta \mathbf{m}\,}\frac{\delta F}{\delta D}\right){\rm
d}^3\mathbf{x}
\\
\nonumber &- \int \left\langle G,\left.\pounds_{\delta F/\delta
\mathbf{m}\,}\frac{\delta K}{\delta G}-\pounds_{\delta H/\delta
\mathbf{m}\,}\frac{\delta F}{\delta G}\right.\right\rangle{\rm d}^3\mathbf{x}
\\
\nonumber &- \int \varphi\left(\pounds_{\delta F/\delta
\mathbf{m}\,}\frac{\delta K}{\delta \varphi}-\pounds_{\delta H/\delta
\mathbf{m}\,}\frac{\delta F}{\delta \varphi}\right){\rm d}^3\mathbf{x}\,{\rm d}n
\\
&+ \int \varphi\left(\left\langle \frac{\partial}{\partial
n}\frac{\delta F}{\delta \varphi},\left(\frac{\delta K}{\delta
G}\right)_{\!\!M} n\right\rangle- \left\langle
\frac{\partial}{\partial n}\frac{\delta K}{\delta
\varphi},\left(\frac{\delta F}{\delta G}\right)_{\!\!M}
n\right\rangle\right){\rm d}^3\mathbf{x}\,{\rm d}n \,.
\label{PB-hybrid-liqXals}
\end{align}
\rem{ %%%%%%%%%%%%%%%%%%%%%%%%%%%%%%%
In turn, the above bracket can be used together with the Hamiltonian
\begin{equation}
H=\frac12\int \frac{\left|M\right|^2}{D}\,\de \mathbf{x}+\frac1{2J}\int\left|G\right|^2\,\de \mathbf{x}+\int\! D\,\mathcal{F}(D^{-1},\varphi,\nabla\varphi)\,\de \mathbf{x}\,,
\end{equation}
where $\mathcal{F}$ corresponds to the hybrid expression of the Frank energy
\begin{multline}
\mathcal{F}=K_1\left|\int\! \boldsymbol{n}\cdot\nabla\varphi\,\de \boldsymbol{n}\right|^{2}+K_2\left|\int\!\varphi(\mathbf{x},\boldsymbol{n}')\ \boldsymbol{n}'\cdot \boldsymbol{n}\times\nabla\varphi(\mathbf{x},\boldsymbol{n})\ \de \boldsymbol{n}' \,\de \boldsymbol{n}\right|^{2}
\\
+K_3\left|\int\!\varphi(\mathbf{x},\boldsymbol{n}')\ \boldsymbol{n}'\times \boldsymbol{n}\times\nabla\varphi(\mathbf{x},\boldsymbol{n})\ \de \boldsymbol{n}' \,\de \boldsymbol{n}\right|^{2}\,,
\end{multline}
to produce the hybrid equations of motion. 
} %%%%%%%%%%%%%%%%%%%%%%%%%%%%%%%
In the incompressible case ($D\equiv 1$) the fluid momentum equals the velocity field $\mathbf{u}$, i.e. $\mathbf{m=u}$. Then, the hybrid equations of motion are
\begin{align*}
&\left(\frac{\partial}{\partial t}+\pounds_{\delta H/\delta\mathbf{u}}\right)\mathbf{u} =-\left\langle G,\nabla\frac{\delta H}{\delta G}\right\rangle+\int\!\frac{\delta H}{\delta\varphi}\,\nabla\varphi\,\de{n}-\nabla\mathsf{p}\,,
\\
&\,\frac{\partial G}{\partial t}+\operatorname{div}\!\left({G}\,\frac{\delta H}{\delta\mathbf{u}}\right)=-\operatorname{ad}^*_{\delta H/\delta G\,}G-\int\!\varphi\,{n} \diamond\frac{\partial}{\partial{n}}\frac{\delta H}{\delta\varphi}\,\de{n}
\\
&\,\frac{\partial \varphi}{\partial t}+\frac{\delta H}{\delta\mathbf{u}} \cdot\!\nabla\varphi=-\frac{\partial}{\partial{n}}\cdot\left(\varphi\left(\frac{\delta H}{\delta G}\right)_M {n} \right)\,,
\end{align*}
\rem{ %%%%%%%%%%%%%%%%%%%%%%%%%%%%%%%%%
A possible explicit expression of this free energy is given by
\begin{multline}
\mathcal{F}=K_1\left(\int\! \boldsymbol{n}\cdot\nabla\varphi\,\de \boldsymbol{n}\right)^{2}+K_3\left|\int\! \boldsymbol{n}\times\nabla\varphi\,\de \boldsymbol{n}\right|^{2}
\\
+\left(K_2-K_3\right)\!\int\!\varphi(\mathbf{x},\boldsymbol{n})\left(\boldsymbol{n}\cdot\int \boldsymbol{n}'\times\nabla\varphi(\mathbf{x},\boldsymbol{n}')\ \de \boldsymbol{n}' \right)^{\!2}\de \boldsymbol{n}
\,.
\end{multline}
} %%%%%%%%%%%%%%%%%%%%%%%%%%%%%%%%%
Notice that, when $M=S^2$ {\color{black}so that $n$ is denoted by the unit vector $\boldsymbol{n}$ and the Yang-Mills charge $G$ by $\mathbf{G}$}, the relation $\mathbf{G\cdot n}=\mathbf{G}\cdot \int\!\boldsymbol{n}\,\varphi\,\de\boldsymbol{n}=0$ is \emph{not} preserved by the hybrid dynamics, {\color{black}when} the free energy $\mathcal{F}$ contains moments of order higher than one. Thus, in this case, molecules that are initially uniaxial {(i.e. satisfying $\mathbf{G}\cdot \int\!\boldsymbol{n}\,\varphi\,\de\boldsymbol{n}=0$)} can change their {\color{black}configuration} as time goes by. This is a typical situation in liquid crystal dynamics. For example, {\color{black}phase transitions (e.g. from uniaxial to biaxial) may occur} in presence of defects, such as disclination lines. 

Also, notice that moments $\int\!\boldsymbol{n}\dots\boldsymbol{n}\,\varphi\,\de\boldsymbol{n}$ of the distribution $\varphi$ can be taken at any order, thereby generating a hierarchy that can be truncated at pleasure by simply assuming that the energy does not involve moments higher than a desired order. Indeed, the equation of the $k$-th moment does not involve higher order moments, as it usually happens in kinetic theory. Rather, the equation of the $k$-th moment involves only moments of the same order $k$. { Thus, the hybrid method proposed here always gives a closed system of equations for an arbitrary multipolar order parameter.}  Of particular interest in liquid crystal dynamics are moment quantities of the type
\begin{equation}\label{tensorOP}
\mathsf{Q}(\mathbf{x},t)=J\int\!\left(\mathbf{I}-\alpha\,\boldsymbol{n}\boldsymbol{n}\right)\varphi(\mathbf{x},\boldsymbol{n},t)\,\de\boldsymbol{n}\,.
\end{equation}
where $J$ and $\alpha$ are constant factors. For example, when $\alpha=1$, the moment $\mathsf{Q}$ is known as \emph{microinertia tensor} in Eringen's micropolar theory \cite{Eringen}, while $\alpha=3$ yields a traceless \emph{alignment tensor} that is used in the Landau-de Gennes theory \cite{deGennes1971}. It is perhaps not surprising that the equation for $\mathsf{Q}(\mathbf{x},t)$ turns out to be identical to that recently derived in \cite{GayBalmazTronci} on a purely geometric basis. {\color{black}The same dynamics for $\mathsf{Q}$ was first found by Volovik in \cite{VoKa}.} Indeed, the standard methods presented at the beginning of this paper can be used to show that the whole Poisson bracket \eqref{PB-hybrid-liqXals} can be written in terms of the alignment tensor $\mathsf{Q}$, thereby obtaining exactly the same Hamiltonian structure as in section 5.1.2 of \cite{GayBalmazTronci}. 
\rem{ %%%%%%%%%%%%%%%%%%%%%%%%%%%%%%%%%%%%%%%%
More particularly, this can be seen upon replacing the variational chain rule relation
\[
\frac{\delta F}{\delta \varphi}=J\operatorname{Tr}\!\left(\left(\mathbf{I}-\alpha\,\boldsymbol{n}\boldsymbol{n}\right)\frac{\delta \bar{F}}{\delta \mathsf{Q}}\right)
\]
in the Poisson bracket \eqref{PB-hybrid-liqXals}.
}     %%%%%%%%%%%%%%%%%%%%%%%%%%%%%%%%%%%%%%%%
{\color{black}Then, the expression of the free energy $F(\mathsf{Q},\nabla\mathsf{Q})$ usually coincides with} the Landau-deGennes free energy \cite{deGePr1993}. 

The dynamics of $\mathsf{Q}$ is particularly important in the case when phase transitions (e.g. from uniaxial to biaxial) may occur in the system. This situation typically occurs in presence of disclination lines, which is an interesting case to study. This is the topic of the next section.

\section{Hybrid description of liquid crystals with disclinations}

When orientational defects are present in liquid crystals, the dynamics of these defects is usually related to an equation of motion for a Yang-Mills magnetic field, which is known as \emph{disclination density}. This is the well known gauge-theory approach to defect dynamics \cite{KaEd,DzVo,VoDz}. In the literature, the dynamics of the disclination density is  given through the dynamics of the corresponding vector potential. In more geometric terms, the vector potential is a \emph{principal connection}, while its covariant derivative is the curvature of the connection. 
%When this covariant derivative vanishes, the connection is trivial and this means that defects are absent. 
Thus, the starting point is the emergence  of a magnetic potential in the fluid model. This can be easily recognized by invoking reduction theory principles \cite{Ho2002, GayBRatiu} to write the field $\mathsf{n}\in C^\infty(\Bbb{R}^3,M)$ in terms of the order parameter group element $\chi\in C^\infty(\Bbb{R}^3,\mathcal{O})$:
\[
\mathsf{n}=
\eta_*\big(\chi\mathsf{n}_0\big)\,,
%\chi(\mathbf{x})\mathsf{n}_0\,,
\]
where $\mathsf{n}_0$ is a (space-independent) fixed reference configuration, typically $\mathsf{n}_0=\left(0,0,1\right)$, and $\eta_*$ is the push-forward of the diffeomorphism $\eta=\exp\!\left(\delta H/\delta \mathbf{m}\right)$ transporting all fluid particles. 
Then, upon using standard properties of the push-forward \cite{MaRa,HoScSt}, simple differentiation yields
\begin{equation}\label{GRAD-n}
\nabla\mathsf{n}=
\eta_*\big(\nabla\chi\,\mathsf{n}_0\big)=\eta_*\big(\!\left(\nabla\chi\chi^{-1}\right)\chi\mathsf{n}_0\big)=\eta_*\!\left(\nabla\chi\chi^{-1}\right)\eta_*\chi\mathsf{n}_0=:{\boldsymbol\gamma}_M\mathsf{n}\,,
\end{equation}
where $\boldsymbol\gamma=\eta_*\big(\left(\nabla\chi\right)\chi^{-1}\big)\in\Omega^1(\Bbb{R}^3,\mathfrak{o})$ is 
%a connection one-form, that is 
a magnetic vector potential. In the general case, one assumes that there exists a vector potential $\gamma$ such that the relation $\nabla\mathsf{n}={\boldsymbol\gamma}_M\mathsf{n}$ is verified at all times. As a result, the gradients of the order parameter field are naturally {\color{black}encoded} in the vector potential, which is a new dynamical variable.

\subsection{Derivation of the Hamiltonian structure}
In the search of a hybrid model describing liquid crystal dynamics in presence of disclination lines, one can proceed by allowing for both magnetic and electric Yang-Mills fields and then operating on the Poisson bracket. One can start by considering the {\color{black}direct-sum bracket
\begin{align}\nonumber
\{F,K\}(\mathbf{m},D,G,\varphi, \mathbf{A, E})=& - \left\{F,K\right\}(\mathbf{A,E})+\{F,K\}(\mathbf{m},D,G,\varphi)  \,,
\end{align}
where the second term on the right hand side is given by \eqref{PB-hybrid-liqXals}. Then, one can} perform the change of variables
\begin{eqnarray*}
G&\mapsto&G-\operatorname{div}^\mathbf{A}\mathbf{E}=:\mathcal{Q}
\\
\mathbf{m}&\mapsto&\mathbf{m}+\mathbf{E}_a\times\operatorname{curl}\mathbf{A}^a-\mathbf{A}^a\operatorname{div}\mathbf{E}_a=:\boldsymbol{{\mathsf{M}}}\,.
\end{eqnarray*}
The momentum shift in the second line was already used for electromagnetic charged fluids by Holm \cite{Ho1986, Ho1987}, who showed how this transformation leads to the Poisson bracket of plasma MHD. The first line is also used here to account for charge neutrality, i.e. $\operatorname{div}^\mathbf{A}\mathbf{E}=0$. Both the above change of variables are given as shifts by a momentum map. Indeed, while the term $-\operatorname{div}^\mathbf{A}\mathbf{E}$ is a momentum map arising from the action of the gauge group (as it appears also in electromagnetic plasmas \cite{MaWeRaScSp}), the quantity $\mathbf{E}_a\times\operatorname{curl}\mathbf{A}^a-\mathbf{A}^a\operatorname{div}\mathbf{E}_a$ is a momentum map arising from the diffeomorphism action {\color{black}on the field variables}. The same quantity also emerges naturally in electromagnetic charged fluids \cite{Ho1986, Ho1987}. This construction allows to apply a { general} result from \cite{KrMa1987} (see proposition 2.2), which {\color{black}generates the resulting Poisson bracket.}
%%%%%%%%%%%%%%%%%%%%%%%%%%%%%%%%%%%%%%%%%%%%%%%%%%%%%%%%%
{\color{black}Then, upon using the neutrality hypothesis
\[
\operatorname{div}^\mathbf{A}\mathbf{E}=0\Rightarrow\mathcal{Q}=G
\,,
\]
the assumption $\delta F/\delta \mathbf{E}=\delta K/\delta \mathbf{E}=0$ yields the final Poisson bracket for
the MHD limit:}
%. When the latter is expressed in terms of $\varphi$, this Poisson structure reads as
\begin{align}\nonumber
\{F,K\}(\mathbf{m},D,G,\rho, {\boldsymbol\gamma})=&  \int \mathbf{m}\cdot \left[\frac{\delta
F}{\delta \mathbf{m}},\frac{\delta K}{\delta \mathbf{m}}\right]_\mathfrak{\!X}{\rm
d}q+\int \left\langle G,\left[\frac{\delta F}{\delta G},\frac{\delta
K}{\delta G}\right]\right\rangle{\rm d}^3\mathbf{x}
\\ \nonumber
&- \int D\left(\pounds_{\delta F/\delta \mathbf{m}\,}\frac{\delta K}{\delta
D}-\pounds_{\delta K/\delta \mathbf{m}\,}\frac{\delta F}{\delta D}\right){\rm
d}q
\\
\nonumber &- \int \left\langle G,\left.\pounds_{\delta F/\delta
\mathbf{m}\,}\frac{\delta K}{\delta G}-\pounds_{\delta K/\delta
\mathbf{m}\,}\frac{\delta F}{\delta G}\right.\right\rangle{\rm d}^3\mathbf{x}
\\
\nonumber &- \int \varphi\left(\pounds_{\delta F/\delta
\mathbf{m}\,}\frac{\delta K}{\delta \varphi}-\pounds_{\delta K/\delta
\mathbf{m}\,}\frac{\delta F}{\delta \varphi}\right){\rm d}^3\mathbf{x}\,{\rm d}n
\\
\nonumber &- \int \left\langle \boldsymbol\gamma,\left.\pounds_{\delta F/\delta
\mathbf{m}\,}\frac{\delta K}{\delta {\boldsymbol\gamma}}-\pounds_{\delta K/\delta
\mathbf{m}\,}\frac{\delta F}{\delta {\boldsymbol\gamma}}\right.\right\rangle{\rm d}^3\mathbf{x}
\\ \nonumber
&+ \int \varphi\left(\left\langle \frac{\partial}{\partial
n}\frac{\delta F}{\delta \varphi},\left(\frac{\delta K}{\delta
G}\right)_{\!\!M} n\right\rangle-\left\langle
\frac{\partial}{\partial n}\frac{\delta K}{\delta
\varphi},\left(\frac{\delta F}{\delta G}\right)_{\!\!M}
n\right\rangle\right){\rm d}^3\mathbf{x}\,{\rm d}n
\\
&+\int\!\left(\left\langle{d^{\boldsymbol\gamma}}\frac{\delta F}{\delta
G},\frac{\delta K}{\delta
\boldsymbol\gamma}\right\rangle-\left\langle{d^{\boldsymbol\gamma}}\frac{\delta K}{\delta
G},\frac{\delta F}{\delta {\boldsymbol\gamma}}\right\rangle\right)\de \mathbf{x}
\label{PB+gamma}
\end{align}
where we have replaced $\boldsymbol{{\mathsf{M}}}$ by $\mathbf{m}$ for simplicity {\color{black}and $\mathbf{A}$ by $\boldsymbol\gamma$ to comply with the notation used in the literature on liquid crystals}. The resulting equations of motion read
\begin{align}\label{Hybrid+defects1}
&\left(\frac{\partial}{\partial t}+\pounds_{\delta H/\delta\mathbf{m}}\right)\mathbf{m} =-D\nabla\frac{\delta H}{\delta D}-\left\langle G,\nabla\frac{\delta H}{\delta G}\right\rangle+\int\!\frac{\delta H}{\delta\varphi}\,\nabla\varphi\,\de{n}+{\boldsymbol\gamma}^a\times\operatorname{curl}\frac{\delta H}{\delta {\boldsymbol\gamma}^a}-{\boldsymbol\gamma}^a\operatorname{div}\frac{\delta H}{\delta {\boldsymbol\gamma}^a}\,,
\\
&\,\frac{\partial G}{\partial t}+\operatorname{div}\!\left({G}\,\frac{\delta H}{\delta\mathbf{m}}\right)=\operatorname{ad}^*_{\delta H/\delta G\,}G-\int\varphi\,{n} \diamond\frac{\partial}{\partial{n}}\frac{\delta H}{\delta\varphi}\,\de{n} -\operatorname{div}^{\boldsymbol\gamma}\frac{\delta H}{\delta {\boldsymbol\gamma}}
\\
&\,\frac{\partial \varphi}{\partial t}+\frac{\delta H}{\delta\mathbf{m}} \cdot\!\nabla\varphi=-\frac{\partial}{\partial{n}}\cdot\left(\varphi\left(\frac{\delta H}{\delta G}\right)_M {n} \right)\,,
\\
&\left(\frac{\partial}{\partial t}+\pounds_{\delta H/\delta\mathbf{m}}\right)\mathbf{{\boldsymbol\gamma}} =-\nabla^{{\boldsymbol\gamma}\,} \frac{\delta H}{\delta G}
\label{Hybrid+defects4}
\end{align}
Here the indexes $a,b,c,\dots$ are Lie algebra indexes. These equations hold for an arbitrary form of the Hamiltonian $H$. Indeed, no specific assumption has been made on the form of the total energy. In the case of liquid crystals, one would hope to use Frank's expression of the free energy, which however still needs to be expressed in terms of the connection $\boldsymbol\gamma$. The explicit form of Frank's energy is presented in the next section.

{
\begin{remark}[Disclinations] Disclinations are defined as discontinuities of the order parameter field $\mathsf{n}$ and the main difficulty involved in the study of their dynamics is that the Frank energy \eqref{FrankEnergy} diverges in the presence of discontinuities. The same problem emerges in the hybrid approach, because of the emergence of the gradient $\nabla\varphi$, which blows up for (spatially) discontinuous distributions $\varphi$. This problem can be avoided by noticing that the relation
\[
\nabla\varphi=\frac{\partial}{\partial{n}}\cdot\Big(\varphi\,{\boldsymbol\gamma}_M {n} \Big)
\]
is preserved by the dynamics, as it can be shown by a tedious computation (or by general principles in reduction theory, see \cite{GBRaTr} and remark \ref{ActionOrdPar}). Then, if one replaces gradients $\nabla\varphi$ according to the above relation, all divergences that appeared in the model are eliminated, as long as all other quantities are smooth. In \cite{GBRaTr}, this fact was used to explore the various relations occurring among different dynamical models for liquid crystals.
\end{remark}
}

\begin{remark}[Kelvin-Noether theorem]
Notice that considerations analogue to those in Section \ref{Hybrid+Density} lead the Kelvin circulation dynamics associated to equation \eqref{Hybrid+defects1}. Indeed, upon repeating analogous steps, one readily obtains the following Kelvin circulation dynamics:
\[
\frac{d}{dt}\oint_\gamma\frac{\mathbf{m}}D\cdot\de\mathbf{x}=-\oint_\gamma\frac1D\left(\left\langle G,\nabla\frac{\delta H}{\delta G}\right\rangle-\int\!\frac{\delta H}{\delta\varphi}\,\nabla\varphi\,\de{n}-\boldsymbol{A}^a\times\operatorname{curl}\frac{\delta H}{\delta \boldsymbol{A}^a}+\boldsymbol{A}^a\operatorname{div}\frac{\delta H}{\delta \boldsymbol{A}^a}\right)\cdot\de\mathbf{x}
\]
where $\gamma$ is a loop moving with the fluid velocity $\mathbf{u}=\delta H/\delta\mathbf{m}$.
\end{remark}
%%%%%%%%%%%%%%%%%%%%%%%%%%%%%%%%%%%%%%%%%%%%%%%%%%%%%%%%%%%%%%

\subsection{The expression of the free energy}
In order to use the dynamical model in the previous section, it is necessary to express Frank's energy in the correct set of variables, including the connection describing defect dynamics. In Eringen's theory of liquid crystals \cite{Eringen}, this quantity is denoted by ${\gamma}$ and it is given by ${\gamma}_j=\partial_j\chi\,\chi^T$, due to the orthogonal property of $\chi\in C^\infty(\Bbb{R}^3,SO(3))${, see equation \eqref{GRAD-n}}. In order to express Frank's energy \eqref{FrankEnergy} in terms of ${\gamma}$, one can use the relation
\[
\partial_j\mathbf{n}={\gamma}_j\,\mathbf{n}=\boldsymbol\gamma_j\times\mathbf{n}\,
\]
where the second equality makes convenient use of the well known hat-map isomorphism $\mathfrak{so}(3)\simeq\Bbb{R}^3$ \cite{MaRa,HoScSt}, so that $\gamma_j=\hat{\boldsymbol\gamma}_j$. { Upon fixing  an appropriate basis $\{e_a\}$ on $\mathfrak{so}(3)\simeq\Bbb{R}^3$, the connection form ${\gamma\in\Omega^1(\Bbb{R}^3)\otimes\Bbb{R}^3}$ can be expressed as $\gamma=\gamma_i^a\,e_a\,\de x^i$. In this context, the bold notation $\boldsymbol\gamma$ that is ordinarily used for vectors in $\Bbb{R}^3$ may become confusing and thus we shall relegate this notation to denote only componentwise quantities of the type $\boldsymbol\gamma_j=\gamma_j^a\,e_a$ or $\boldsymbol\gamma^a=\gamma^a_i\,\de x^i$.} For example, one computes
\begin{align*}
\mathbf{n}\cdot\nabla\times\mathbf{n}=&\ n_c\epsilon_{cjd}\partial_j n_d
\\
=&\
n_c\epsilon_{cjd}\epsilon_{dab}\gamma_{aj}n_b
\\
=&\
-n_cn_b\left(\epsilon_{dcj}\epsilon_{dab}\right)\gamma_{aj}
\\
=&\
-n_cn_b\left(\delta_{ca}\delta_{jb}-\delta_{cb}\delta_{ja}\right)\gamma_{aj}
\\
=&\
-n_an_j\gamma_{aj}+\gamma_a^a
\\
=&\
-\operatorname{Tr}\!\left(\mathbf{n}\mathbf{n}\gamma\right)+\operatorname{Tr}\gamma
\,.
\end{align*}
By proceeding analogously, one has
\begin{align*}
\left(\mathbf{n}\times\nabla\times\mathbf{n}\right)_i=&\ \epsilon_{ijk}n_j\epsilon_{kab}\partial_an_b
\\
=&\ 
-\epsilon_{kij}\epsilon_{kab}\,n_j \epsilon_{bcd}\gamma_{ca}n_d
\\
=&\ 
-\left(\delta_{ia}\delta_{jb}-\delta_{ib}\delta_{ja}\right)n_j\epsilon_{bcd}\gamma_{ca}n_d
\\
=&\ 
-\left(\epsilon_{jcd}\gamma_{ci}
+\epsilon_{icd}\gamma_{cj}\right)n_dn_j
\\
=&\ 
-\epsilon_{icd}\,n_d\gamma_{cj}n_j
\\
=&\ 
{\left(\mathbf{n}\times\left(\boldsymbol\gamma_j{n}_j\right)\right)_i}
\,,
\end{align*}
and also
\begin{align*}
\left(\operatorname{div}\mathbf{n}\right)^2=&\
\left(\partial_in_i\right)^2
=
\left(\epsilon_{iab}\gamma_{ai}n_b\right)^2
=
\left(\vec{\boldsymbol{\gamma }}\cdot\mathbf{n}\right)^2
\,,
\end{align*}
where we have defined the operator
${\vec{\eta}}_i:=\epsilon_{ijk}\eta_{jk}$, for any $3\times 3$ matrix $\eta_{jk}$. Then, since
\[
\left|\mathbf{n}\times\left(\boldsymbol\gamma_j{n}_j\right)\right|^2=\left|\boldsymbol\gamma_j{n}_j\right|^2-\left(\mathbf{n}\cdot\left(\boldsymbol\gamma_j{n}_j\right)\right)^2=\operatorname{Tr}\!\left(\mathbf{n}\mathbf{n}\gamma^T\gamma\right)-\operatorname{Tr}^2\!\left(\mathbf{n}\mathbf{n}\gamma\right),
\]
we obtain the following alternative expression for the Frank free energy:
\begin{multline}\label{Franken+gamma}
F(\mathbf{n},\gamma)=K_1\left(\vec{\boldsymbol{\gamma }}\cdot\mathbf{n}\right)^2+\big(\!\operatorname{Tr}\!\left(\mathbf{n}\mathbf{n}\gamma\right)-\operatorname{Tr}\gamma\big)\left(K_2\big(\!\operatorname{Tr}\!\left(\mathbf{n}\mathbf{n}\gamma\right)-\operatorname{Tr}\gamma\big)+K_2'\right)
\\
+K_3\left(\operatorname{Tr}\!\left(\mathbf{n}\mathbf{n}\gamma^T\gamma\right)-\operatorname{Tr}^2\!\left(\mathbf{n}\mathbf{n}\gamma\right)\right)
\end{multline}
Therefore, a possible hybrid (quadratic) expression for the Frank free energy is given by
\begin{multline}\label{FreeEnergy+gamma}
\mathcal{F}(\varphi,\gamma)=K_1\iint\!\varphi(\mathbf{x},\boldsymbol{n})\varphi(\mathbf{x},\boldsymbol{n}')\left(\vec{\boldsymbol{\gamma }}\cdot\boldsymbol{n}\right)\left(\vec{\boldsymbol{\gamma }}\cdot\boldsymbol{n}'\right)\left(\boldsymbol{n}\cdot\boldsymbol{n}'\right)\de\boldsymbol{n}'\,\de\boldsymbol{n}
\\
+K_{2\,}
\iint\!\varphi(\mathbf{x},\boldsymbol{n})\varphi(\mathbf{x},\boldsymbol{n}')\big(\!\operatorname{Tr}\!\left(\boldsymbol{n}\boldsymbol{n}\gamma\right)-\operatorname{Tr}\gamma\big)\big(\!\operatorname{Tr}\!\left(\boldsymbol{n}'\boldsymbol{n}'\gamma\right)-\operatorname{Tr}\gamma\big)\de\boldsymbol{n}'\,\de\boldsymbol{n}
\\
+\frac12K_2'
\iint\!\varphi(\mathbf{x},\boldsymbol{n})\varphi(\mathbf{x},\boldsymbol{n}')\left(\boldsymbol{n}\cdot\boldsymbol{n}'\right)\Big(\!\operatorname{Tr}\!\left(\boldsymbol{n}\boldsymbol{n}'\gamma\right)+\operatorname{Tr}\!\left(\boldsymbol{n}'\boldsymbol{n}\gamma\right)-2\left(\boldsymbol{n}\cdot\boldsymbol{n}'\right)\operatorname{Tr}\gamma\Big)\de\boldsymbol{n}'\,\de\boldsymbol{n}
\\
+K_3\iint\!\varphi(\mathbf{x},\boldsymbol{n})\varphi(\mathbf{x},\boldsymbol{n}')\Big(\!\left(\boldsymbol{n}\cdot\boldsymbol{n}'\right)\operatorname{Tr}\!\left(\boldsymbol{n} \boldsymbol{n}'\gamma^T\gamma\right)-\operatorname{Tr}\!\left(\boldsymbol{n} \boldsymbol{n}'\gamma\right)\operatorname{Tr}\!\left(\boldsymbol{n}' \boldsymbol{n}\gamma\right)\!\Big)\de\boldsymbol{n}'\,\de\boldsymbol{n}\,,
\end{multline}
which evidently reduces to \eqref{Franken+gamma} upon restricting to the singular expression ${\varphi=\delta(\boldsymbol{n}-\mathbf{n}(\mathbf{x}))}$ (with $\mathbf{n}(\mathbf{x})\in S^2$).  The {\color{black}Mauer-Saupe} factor $\boldsymbol{n}\cdot\boldsymbol{n}'$ has been introduced to make expression \eqref{FreeEnergy+gamma} symmetric with respect to permutations $\boldsymbol{n}\leftrightarrow\boldsymbol{n}'$ and $\Bbb{Z}_2-$reflections $\boldsymbol{n}\to-\boldsymbol{n}$ (or $\boldsymbol{n}'\to-\boldsymbol{n}'$). 
Notice how the free energy \eqref{FreeEnergy+gamma}  is completely determined by $\gamma$ and the quadrupole moment field $\int\!\boldsymbol{n} \boldsymbol{n}\,\varphi\,\de \boldsymbol{n}$. However,  other forms of the hybrid free energy may also involve higher-order moments,
\rem{ %%%%%%%%%%%%%%%%%%%%%%%%%%%%%%%%%%%%%%%%%%%%%%%%%%%%%%%%
 through the term
\[
\iint\!\varphi(\mathbf{x},\boldsymbol{n})\operatorname{Tr}^2\!\left(\boldsymbol{n}\boldsymbol{n}{\gamma}\right)\,\de\boldsymbol{n}\,.
\]
Higher order moments of this type also
} %%%%%%%%%%%%%%%%%%%%%%%%%%%%%%%%%%%%%%%%%%%%%%%%%%%%%%%% 
{\color{black}similarly to what happens} in the derivation of the celebrated Doi theory \cite{DoEd1988}.
\begin{remark}[Cubatic liquid crystals]
 Multipolar expressions {\color{black}emerge also in the description} of cubatic liquid crystal phases \cite{VeFr,DuDeMaWi,BaStTo}, although expressing {\color{black}their free energy} remains a difficult task. For example, in \cite{DuDeMaWi} the order parameter of cubatic phases is derived by making use of the 4-tensor
\begin{multline*}
\mathcal{Q}_{\alpha\beta\gamma\mu}=\frac{35}8\!\int\!\varphi\, n_\alpha n_\beta n_\gamma n_\mu\, \de n+\frac18\left(\delta_{\alpha\beta}\delta_{\gamma\mu}+\delta_{\alpha\gamma}\delta_{\beta\mu}+\delta_{\alpha\mu}\delta_{\beta\gamma}\right)
\\
-\frac58\!\int\!\varphi\left(
n_\alpha n_\beta \delta_{\gamma\mu}
+
n_\alpha n_\gamma \delta_{\beta\mu}
+
n_\alpha n_\mu\delta_{\beta\gamma}
+
n_\beta n_\gamma\delta_{\alpha\mu}
+
n_\beta n_\mu\delta_{\alpha\gamma}
+
n_\gamma n_\mu\delta_{\alpha\beta}
\right)\de n\,,
\end{multline*}
{\color{black}whose dynamics can be derived from the hybrid approach by simply taking moments of the $\varphi$-equation, upon adding the appropriate terms to the expression of the free energy $\mathcal{F}(\varphi,\boldsymbol\gamma)$.}
\end{remark}
At this point, one would like to use the energy \eqref{FreeEnergy+gamma} to formulate an explicit model for the dynamics of liquid crystals in the presence of disclinations. However, 
%when these orientational defects are considered, an explicit expression of the total energy cannot be found in the general case. Indeed, 
the presence of defects is responsible for molecular { configuration} changes, thereby producing a varying microinertia tensor of the type \eqref{tensorOP}, which {\color{black}modifies the expression of the total energy.}
\rem{ %%%%%%%%%%%%%%%%%%%%%%%%%%%%%%%%%%%%%%%%%%%%%%%%%%%%%%%%%%%%%%%%%%%%%
is not invertible in the general case. Thus, it is common to derive explicit equations of liquid crystals from Hamilton's principle \cite{Ho2002,GayBRatiu}, so that the whole dynamics is naturally encoded in a certain Lagrangian. Nevertheless, in order to stay with the Hamiltonian construction, the next section presents a formal expression of the total energy, which will produce the final version of the { correct} hybrid equations for liquid crystals. 
}    %%%%%%%%%%%%%%%%%%%%%%%%%%%%%%%%%%%%%%%%%%%%%%%%%%%%%%%%%%%%%%

\subsection{Hybrid equations and relations to other models}
Upon restricting to the incompressible case, a possible hybrid expression of the Hamiltonian for liquid crystal dynamics is given by the functional 
\begin{equation}
H=\frac12\int \!\left|\mathbf{u}(\mathbf{x})\right|^2\de \mathbf{x}+\frac12\!\int\! \boldsymbol{G}(\mathbf{x})\cdot\mathsf{Q}(\varphi)^{-1}\boldsymbol{G}(\mathbf{x})\,\de\boldsymbol{n}\,\de \mathbf{x}+\int\!\mathcal{F}(\varphi,\gamma)\,\de\boldsymbol{n}\,\de \mathbf{x}\,,
\end{equation}
where the microinertia  tensor $\mathsf{Q}(\varphi)=J\int\varphi\left(\mathbf{I}-\boldsymbol{n}\boldsymbol{n}\right)\de\boldsymbol{n}$ evidently appears in the rotational energy, in analogy with rigid body dynamics. Notice that the above Hamiltonian assumes that the microinertia tensor is invertible at all times, while this is not necessarily true in the general case. At this point, inserting the above Hamiltonian in equations \eqref{Hybrid+defects1}-\eqref{Hybrid+defects4} yields
\begin{align*}
&\frac{\partial\mathbf{u}}{\partial t}+\left(\mathbf{u}\cdot\nabla\right)\mathbf{u}-\nabla\mathsf{p} =\int\!\frac{\delta \mathcal{F}}{\delta\varphi}\,\nabla\varphi\,\de\boldsymbol{n}+\boldsymbol{\gamma}^a\times\operatorname{curl}\frac{\partial \mathcal{F}}{\partial \boldsymbol{\gamma}^a}\,-\boldsymbol{\gamma}^a\operatorname{div}\frac{\partial \mathcal{F}}{\partial \boldsymbol{\gamma}^a}\,,
\\
&
\frac{\partial\boldsymbol{G}}{\partial t}+(\mathbf{u}\cdot\nabla)\boldsymbol{G}+\boldsymbol{G}\times\!\left(\mathsf{Q}^{-1}\boldsymbol{G}\right)
=-\frac{J}2\!\int\!\Big(\mathsf{Q}^{-1}\boldsymbol{G}\cdot (\mathbf{I}-\boldsymbol{n}\boldsymbol{n})\mathsf{Q}^{-1}\boldsymbol{G}\Big)\boldsymbol{n}\times\frac{\partial\varphi}{\partial\boldsymbol{n}}\,\de\boldsymbol{n}
\\
&\hspace{7.5cm}
-\!\int\!\varphi\,\boldsymbol{n} \times\frac{\partial}{\partial\boldsymbol{n}}\frac{\delta \mathcal{F}}{\delta\varphi}\,\de\boldsymbol{n} -\operatorname{div}^{{\gamma}}\frac{\partial \mathcal{F}}{\partial{\gamma}}
\\
&\,\frac{\partial \varphi}{\partial t}+\mathbf{u} \cdot\!\nabla\varphi=\left(\mathsf{Q}^{-1}\boldsymbol{G}\right)\cdot\boldsymbol{n}\times\frac{\partial\varphi}{\partial\boldsymbol{n}}\,,
\\
&\partial_t{\boldsymbol\gamma_i}+(\partial_iu^j)\boldsymbol\gamma_j+(\mathbf{u}\cdot\nabla){\boldsymbol{\gamma}_i}=-\partial_i\!\left(\mathsf{Q}(\varphi)^{-1}\boldsymbol{G}\right)+\boldsymbol{\gamma}_i\times\!\left(\mathsf{Q}^{-1}\boldsymbol{G}\right)
\end{align*}
In order to get rid of the inverse matrix $\mathsf{Q}(\varphi)^{-1}$, we may introduce the angular velocity variable $\boldsymbol\nu$ in analogy with rigid body motion, such that $\boldsymbol{G}=\mathsf{Q}(\varphi)\boldsymbol\nu$.  As a result, we have the final form of the hybrid equations:
\begin{align}\label{FinalHybrid1}
&\frac{\partial\mathbf{u}}{\partial t}+\left(\mathbf{u}\cdot\nabla\right)\mathbf{u}-\nabla\mathsf{p} =-\int\!\frac{\delta \mathcal{F}}{\delta\varphi}\,\nabla\varphi\,\de{n}-\boldsymbol{\gamma}^a\times\operatorname{curl}\frac{\partial \mathcal{F}}{\partial \boldsymbol{\gamma}^a}+\boldsymbol{\gamma}^a\operatorname{div}\frac{\partial \mathcal{F}}{\partial \boldsymbol{\gamma}^a}\,,
\\
&
\mathsf{Q}(\varphi)\!\left(\frac{\partial\boldsymbol\nu}{\partial t}+(\mathbf{u}\cdot\nabla)\boldsymbol\nu\right)-\mathsf{Q}(\varphi)\,\boldsymbol\nu\times\boldsymbol\nu
=-\int\!\varphi\,\boldsymbol{n} \times\frac{\partial}{\partial\boldsymbol{n}}\frac{\delta \mathcal{F}}{\delta\varphi}\,\de\boldsymbol{n} -\operatorname{div}^{{\gamma}}\frac{\partial \mathcal{F}}{\partial{\gamma}}
\\
&\,\frac{\partial \varphi}{\partial t}+\mathbf{u} \cdot\!\nabla\varphi=\boldsymbol\nu\cdot\boldsymbol{n}\times\frac{\partial\varphi}{\partial\boldsymbol{n}}\,,
\label{FinalHybrid3}
\\
&\partial_t{\boldsymbol\gamma_i}+(\partial_iu^j)\boldsymbol\gamma_j+(\mathbf{u}\cdot\nabla){\boldsymbol{\gamma}_i}=-\partial_i\boldsymbol\nu+\boldsymbol{\gamma}_i\times\boldsymbol\nu
\label{FinalHybrid4}
\end{align}
which follow upon making use of the relations
\begin{align*}
&J\int\!\left(\frac{\partial\varphi}{\partial t}+(\mathbf{u}\cdot\nabla)\varphi\right)\!\left(\mathbf{I}-\boldsymbol{n}\boldsymbol{n}\right)\boldsymbol\nu\,\de n\,=\boldsymbol\nu\times\mathsf{Q}\boldsymbol\nu
\\
&\frac{J}2\int\!\varphi\left(\boldsymbol{n}\times\frac{\partial}{\partial\boldsymbol{n}}\right)\big(\boldsymbol{\nu}\cdot\left(\mathbf{I}-\boldsymbol{n}\boldsymbol{n}\right)\boldsymbol{\nu}\big)\,\de{n}=-\boldsymbol\nu\times\mathsf{Q}\boldsymbol\nu\,,
\end{align*}
obtained {\color{black}from} equation \eqref{FinalHybrid3} {\color{black}by standard tensor index computations}.

At this point, one may wonder how equations \eqref{FinalHybrid1}-\eqref{FinalHybrid4} are related to ordinary models for complex fluids, under the substitution $\varphi=\delta(\boldsymbol{n}-\mathbf{n}(\mathbf{x},t))$, with $\mathbf{n}(\mathbf{x},t)\in S^2$. {\color{black}Indeed}, since the Free energy was derived from the Frank energy upon using $\partial_i\mathbf{n}=\boldsymbol{\gamma}_i\times\mathbf{n}$, the above hybrid model is compatible with Ericksen-Leslie dynamics \cite{GBRaTr}. In particular, if one assumes that no defects are present at the initial time (i.e. $\gamma(\mathbf{x},0)=0$) and upon restricting to rod-like molecules (i.e. $\boldsymbol\nu\cdot\mathbf{n}=0$), this compatibility is ensured by geometric symmetry arguments \cite{GayBRatiu,GBRaTr,GayBalmazTronci, Ho2002}.

 An interesting observation also arises when the hybrid equations are written in terms of the moments $\int\!\boldsymbol{n}\boldsymbol{n}\dots\boldsymbol{n}\,\varphi(\mathbf{x},\boldsymbol{n},t)\,\de\boldsymbol{n}$. More particularly, since the free energy  \eqref{FreeEnergy+gamma} depends only on second-order moments,
 %both the quantity $\mathbf{n}(\mathbf{x})\mathbf{n}(\mathbf{x})=\int\boldsymbol{n}\,\boldsymbol{n}'\,\varphi(\mathbf{x},\boldsymbol{n})\,\varphi(\mathbf{x},\boldsymbol{n}')\,\de \boldsymbol{n}'\,\de \boldsymbol{n}$ and 
 %As a result, one needs to consider \emph{both first and second order moments} of $\varphi$. However, let us assume for now that correlation properties are such that the total energy can be completely expressed in terms of the alignment tensor $\mathsf{Q}$. 
then the total energy can be completely expressed in terms of the symmetric microinertia tensor \eqref{tensorOP} (with $\alpha=1$). In this way,
the Poisson bracket \eqref{PB+gamma} can be transformed into the Poisson bracket of Eringen's equations for micropolar media \cite{Eringen} (see section 8.11.2 in \cite{GayBRatiu}). Therefore, the hybrid equations of motion recover Eringen's description of micropolar liquid crystals. {\color{black}Notice that the same bracket structure involving the tensor \eqref{tensorOP} appeared previously in \cite{VoKa}.}

%As we already observed, other moment orders can be easily considered in the above description. For example, the first order moment $\mathbf{n}=\int\!\boldsymbol{n}\varphi\,\de\boldsymbol{n}$ may also appear in the free energy. Then the presence of $\gamma$ in the resulting Hamiltonian structure extends the theory of Lhuillier-Rey for ordered micropolar media  \cite{LhuRe} (see also equations (8.70)-(8.71) in \cite{GayBRatiu}) to the case of non-zero disclination density. 

Higher moment orders may also appear in the expression of the free energy. Then, the moment hierarchy can be truncated at the highest order and the resulting theory generalizes ordinary fluid models of liquid crystal dynamics. Indeed,  {\color{black}upon} taking the $k$-th order moment of equation \eqref{FinalHybrid3}, the resulting moment equation does \emph{not} involve any moment of order higher than $k$, differently from other usual kinetic theory approaches. This result may be used, for example, for defect dynamics in cubatic liquid crystals, whose order parameter can be defined as a combination of higher order moments, as shown in \cite{DuDeMaWi}. This interesting possibility will be explored elsewhere.

{
\begin{remark}[Differences with Doi's theory]
Besides considering dominant inertial effects and neglecting dissipation, the present approach differs from Doi's theory of polymeric fluids \cite{DoEd1988} also because it encodes disclination dynamics. This has always been an outstanding task in  Smoluchowski-like models, since it is not clear how disclinations can be incorporated into the framework of kinetic theory. The present hybrid approach solves this problem in the energy-conserving case, in which moment equations close naturally at any order.
\end{remark}
}

\begin{remark}[The action of the order parameter gauge group]\label{ActionOrdPar}
Notice that the construction above can be easily generalized to an arbitrary order parameter group $\mathcal{O}$ and it is geometrically equivalent to the usual geometric construction of complex fluid dynamics \cite{Ho2002,GayBRatiu}, including the affine action of the gauge group $\mathcal{F}(\Bbb{R}^3,\mathcal{O})$ on the space $\Omega^1(\Bbb{R}^3,\mathfrak{o})$ of connection one-forms. Indeed, replacing the order parameter field ${\sf n}\in\mathcal{F}(\Bbb{R}^3,M)$ by the distribution-valued function $\varphi\in\mathcal{F}(\Bbb{R}^3,\mathrm{Den}(M))$ has the only result of changing the action of the gauge group $\mathcal{F}(\Bbb{R}^3,\mathcal{O})$ on the respective quantity. For example, while $\mathcal{F}(\Bbb{R}^3,SO(3))$ acts on $\bf n(x)$ by matrix multiplication (i.e. $\mathbf{n}\mapsto\chi\mathbf{n}$, with $\chi\in\mathcal{F}(\Bbb{R}^3,SO(3))$), the gauge action on $\varphi(\mathbf{x},\boldsymbol{n})$ is  $\varphi(\mathbf{x},\boldsymbol{n})\mapsto\varphi(\mathbf{x},\chi^{-1}\boldsymbol{n})$.
\end{remark}

\section{Conclusions}
This paper has introduced a new approach to complex fluids with multipolar order in which the fluid transports the statistical information that is encoded in a position-dependent probability distribution. In the context of Yang-Mills plasmas, this approach arises as a cold-plasma-like closure of the underlying Vlasov kinetic equation, which possesses momentum map properties. This closure involves moments belonging to different hierarchies thereby producing a hybrid description of the YM plasma. The resulting equations of motion possess a natural Hamiltonian structure, which in turn possesses a whole family of Casimir invariants. As a first application, the example of ferromagnetic hydrodynamics with quadrupolar interactions has been presented and compared with previous results in the literature \cite{Isayev}.

The second part of the paper extended the hybrid formulation of complex fluids to systems exhibiting symmetry breaking. Again, starting from a microscopic kinetic approach, a hybrid moment closure has been presented, which involves a statistical distribution on the order parameter space. The case of broken symmetry presents some difficulties that where overcome by encoding the gradients of the order parameter field in a new variable that accounts for defect dynamics. Motivated by cubatic liquid crystal phases, the dynamics of liquid crystals was taken as the guiding example and a hybrid expression of the Frank free energy was presented. The resulting hybrid liquid crystal model is compatible with Ericksen-Leslie dynamics and it recovers Eringen's formulation of micropolar liquid crystals \cite{Eringen} when the total energy can be expressed in terms of second order moments. {\color{black}The same hybrid model also recovers equations 25 in \cite{VoKa}.} The possibility of higher order moments appearing in the free energy generalizes the theory to account for the higher correlation effects that appear in cubatic phases \cite{VeFr}. 

\subsubsection*{Acknowledgments}
The author is greatly indebted with Fran\c{c}ois Gay-Balmaz and Tudor Ratiu for several stimulating discussions on defect dynamics and its related affine action formulation. Moreover, the author is grateful to Darryl Holm and Vakhtang Putkaradze for illuminating conversations and for their hospitality at Imperial College London and Colorado State University, Fort Collins, where part of this work was carried out. Motivating conversations with Giovanni De Matteis are also greatly acknowledged.

\bigskip

\appendix

\section{A generalization of defects}
In many situations, the elastic energy of complex fluids depends on the gradients of elements $O\in\mathcal{F}(\Bbb{R}^3,\mathcal{O})$ in the gauge group. After symmetry is taken into account, this leads to the introduction of a connection one form $\boldsymbol\gamma(\mathbf{x}) =O^{-1}\nabla O\in\Omega^{1}(\Bbb{R}^3,\mathfrak{o})$, which is also known as \emph{wryness tensor} in the context of liquid crystal dynamics \cite{Eringen}. 
For example, when $\mathcal{O}=SO(3)$, the quantity $\boldsymbol\gamma$ represents orientational defects, such as disclinations whose density is given by the associated curvature $\mathbf{B}=d^{\boldsymbol\gamma}\boldsymbol\gamma$ {\cite{DzVo,VoDz}}. A famous example of how this quantity appears in complex fluids is found in the dynamics of spin glasses {\cite{DzVo,VoDz,HoKu1988}}.

Given the similarity with the gauge approach to complex fluids (see remark \ref{Rem-LPHybrid}), the hybrid construction can be extended to include a connection taking values in the Poisson algebra $C^\infty(\mathfrak{o}^*)$. In this context, the gauge group can be considered to take values in the infinite dimensional (sub)group of canonical transformations $\operatorname{Diff}_{\rm can}(T^*\mathcal{O})$, so that at every point $\mathbf{x}$ in physical space there is an associated canonical transformation $g(\mathbf{x})$ that is responsible for the statistical micromotion. Then, if gradients of the type $\nabla g$  appear in the total energy of the system, one is led to the introduction of the one form 
\[
\boldsymbol\vartheta(\mathbf{x},\sigma)=g^{-1}\nabla g\in \Omega^{1}(\Bbb{R}^3,C^\infty(\mathfrak{o}^*))
\,. 
\]
This quantity is completely absent in soft matter modeling and emerges in the present context for the first time. Its dynamics can be easily derived by applying the affine reduction presented in \cite{GayBRatiu}, so that the Lie-Poisson bracket
\begin{multline*}
\{F,K\}(\mathbf{m},\rho,\boldsymbol\vartheta)= \int \mathbf{m}\cdot \left[\frac{\delta F}{\delta
\mathbf{m}},\frac{\delta K}{\delta \mathbf{m}}\right]_\mathfrak{\!X}{\rm d}^3\mathbf{x}+\int
\rho\,\left\langle\sigma,\left[\frac{\partial}{\partial\sigma}\frac{\delta
F}{\delta \rho},\frac{\partial}{\partial\sigma}\frac{\delta
K}{\delta \rho}\right]\right\rangle{\rm d}^3\mathbf{x}\,{\rm d}\sigma
\\
- \int \rho\left(\pounds_{\delta F/\delta \mathbf{m}\,}\frac{\delta K}{\delta
\rho}-\pounds_{\delta K/\delta \mathbf{m}\,}\frac{\delta F}{\delta
\rho}\right)\!{\rm d}^3\mathbf{x}\,{\rm d}\sigma
-
\int {\boldsymbol\vartheta}\cdot\left(\pounds_{\delta F/\delta \mathbf{m}\,}\frac{\delta K}{\delta
{\boldsymbol\vartheta}}-\pounds_{\delta K/\delta \mathbf{m}\,}\frac{\delta F}{\delta
{\boldsymbol\vartheta}}\right){\rm d}^3\mathbf{x}\,{\rm d}\sigma
\\
+\int\!\frac{\delta K}{\delta
{\boldsymbol\vartheta}}\cdot\left(\nabla\frac{\delta F}{\delta
\rho}-\left\langle\sigma,\left[\frac{\partial\boldsymbol\vartheta}{\partial\sigma},\frac{\partial}{\partial\sigma}\frac{\delta F}{\delta\rho}\right]\right\rangle\right)\!\de^3 \mathbf{x}\,\de\sigma
-\int\!\frac{\delta F}{\delta
{\boldsymbol\vartheta}}\cdot\left(\nabla\frac{\delta K}{\delta
\rho}-\left\langle\sigma,\left[\frac{\partial\boldsymbol\vartheta}{\partial\sigma},\frac{\partial}{\partial\sigma}\frac{\delta K}{\delta\rho}\right]\right\rangle\right)\!\de^3 \mathbf{x}\,\de\sigma
%\label{PB-rho&m&vartheta}
\end{multline*}
produces the Lie-Poisson equations
\begin{align*}\nonumber
\left(\frac{\partial}{\partial t}+\pounds_{\delta H/\delta \mathbf{m}}\right)\mathbf{m}&=-\int\!\left(\rho\,\nabla \frac{\delta H}{\delta \rho}+\nabla{\boldsymbol\vartheta}^{T}\cdot\frac{\delta H}{\delta \boldsymbol\vartheta}\right)\de\sigma-\operatorname{div}\!\int\!\frac{\delta H}{\delta \boldsymbol\vartheta} \boldsymbol\vartheta\,\de\sigma
\\
\frac{\partial\rho}{\partial t}+{\rm div\!}\left(\rho\,\frac{\delta
H}{\delta \mathbf{m}}\right)&=-\left\langle\sigma,\left[\frac{\partial
\rho}{\partial \sigma},\frac{\partial}{\partial \sigma}\frac{\delta
H}{\delta \rho}\right]\right\rangle -\operatorname{div}\frac{\delta H}{\delta \boldsymbol\vartheta}-\left\langle\sigma,\left[\frac{\partial
{\vartheta_i}}{\partial \sigma},\frac{\partial}{\partial \sigma}\frac{\delta
H}{\delta \vartheta_i}\right]\right\rangle
\\
\left(\frac{\partial}{\partial t}+\pounds_{\delta H/\delta \mathbf{m}}\right){\boldsymbol\vartheta}&=\nabla\frac{\delta
H}{\delta \rho}-\left\langle\sigma,\left[\frac{\partial
{\boldsymbol\vartheta}}{\partial \sigma},\frac{\partial}{\partial \sigma}\frac{\delta
H}{\delta \rho}\right]\right\rangle 
\,,
%\label{newmodel+defects}
\end{align*}
where summation over repeated indexes has been used. In the absence of fluid flow (i.e. $\mathbf{m}=0$) and in one spatial dimension, the last two equations above  belong to a class of systems whose integrable cases were recently studied in \cite{HoIvPe2011}.

In the context of complex fluids, the connection one form $\boldsymbol\vartheta$ encodes defects associated to the phase space dynamics of the order parameter and its physical implications are yet unknown. Indeed, to the author's knowledge, no Hamiltonian $H(\mathbf{m},\rho,\boldsymbol\vartheta)$ is available in the physics literature. As a conclusive remark, we notice that if such a Hamiltonian depends on $\rho$ only through the first moment $G=\int\!\sigma\,\rho\,\de\sigma$ and $\boldsymbol\vartheta$ is linear in $\sigma$ (i.e. $\boldsymbol\vartheta(\mathbf{x},\sigma)=\left\langle\sigma,\boldsymbol{\gamma}(\mathbf{x})\right\rangle$), then the above dynamics reduces precisely to the Hamiltonian structure of Yang-Mills MHD \cite{HoKu1988,HoKu1984}, with YM magnetic field $\boldsymbol{B}=d^{\boldsymbol{\gamma}}\boldsymbol{\gamma}$ (also known as \emph{disclination density}).

\end{document}